\documentclass[11pt]{article}
\usepackage[OT1]{fontenc}
\usepackage[utf8]{inputenc}
\usepackage[margin=1in]{geometry}
\usepackage{graphicx}
\usepackage{algorithm}
\usepackage{algorithmic}
\usepackage{amsmath}
\usepackage{amsfonts}
\usepackage{comment} 
\usepackage{appendix}
\usepackage{subfigure}
\usepackage[usenames,dvipsnames]{xcolor}
\usepackage{lipsum}  
\usepackage{tikz}
\usepackage{amsthm}
\usepackage[affil-it]{authblk}
\usepackage{url}

\usepackage{setspace}

\usepackage{soul}
\usepackage{times}

\newcommand\nst[1]{{\color{black}#1}} 

\newcommand{\beginsupplement}{%
    \clearpage
    \setcounter{section}{0}%
    \renewcommand{\thesection}{S\arabic{section}}%
    \setcounter{equation}{0}%
    \renewcommand{\theequation}{S.\arabic{equation}}%
    \setcounter{figure}{0}%
    \renewcommand{\thefigure}{S\arabic{figure}}%
    \setcounter{subsection}{0}%
    \renewcommand{\thesubsection}{S\arabic{subsection}}%
}

\title{Oja's plasticity rule overcomes challenges of training neural networks under biological constraints}

\author[1,2,*]{Navid Shervani-Tabar}
\author[1,3]{Marzieh Alireza Mirhoseini}
\author[1,4]{Robert Rosenbaum}
\affil[1]{Department of Applied and Computational Mathematics and Statistics, University of Notre Dame, Notre Dame, IN 46556, USA}
\affil[2]{The Picower Institute for Learning and Memory, Massachusetts Institute of Technology, Cambridge, MA 02139, USA}
\affil[3]{Department of Earth, Atmospheric, and Planetary Sciences, Massachusetts Institute of Technology, Cambridge, MA 02139, USA}
\affil[4]{Department of Biological Sciences, University of Notre Dame, Notre Dame, IN 46556, USA}
\affil[*]{Corresponding author: nshervan@nd.edu}
\date{\today}

\begin{document}
\maketitle    

\begin{abstract}
Deep neural networks have achieved impressive performance through carefully engineered training strategies. Nonetheless, such methods lack parallels in biological neural circuits, relying heavily on non-local credit assignment, precise initialization, normalization layers, batch processing, and large datasets. Biologically plausible plasticity rules, such as random feedback alignment, often suffer from instability and unbounded weight growth without these engineered methods, while Hebbian-type schemes fail to provide goal-oriented credit. In this study, we demonstrate that incorporating Oja’s plasticity rule into error-driven training yields stable, efficient learning in feedforward and recurrent architectures, obviating the need for carefully engineered tricks. Our results show that Oja’s rule preserves richer activation subspaces, mitigates exploding or vanishing signals, and improves short-term memory in recurrent networks. Notably, meta-learned local plasticity rules incorporating Oja’s principle not only match but surpass standard backpropagation in data-scarce regimes. These findings reveal a biologically grounded pathway bridging engineered deep networks and plausible synaptic mechanisms.
\end{abstract}

\section*{}

The practical techniques required to train deep artificial neural networks (DNNs) differ markedly from known mechanisms of synaptic plasticity. In the late 1980s, attempts to construct networks from biologically grounded principles were hindered by fundamental obstacles: non-local weight updates and problems with forward-backward weight alignment~\cite{grossberg1987competitive, kolen1994backpropagation}. While the biologically focused methods stagnated, the 2010s saw a renaissance in artificial networks fueled by the maturation of backpropagation-based methods and critical engineering breakthroughs. These included carefully designed weight-initialization strategies~\cite{glorot2010understanding, he2015delving, arjovsky2016unitary}, advanced gradient-based optimizers~\cite{hinton2012rmsprop, kingma2014adam, pascanu2013difficulty}, and normalization layers~\cite{ioffe2015batch, salimans2016weight, ba2016layer}. These engineering innovations helped modern DNNs overcome instability in activations, vanishing and exploding gradients, and sensitivity to hyperparameters, reaching astonishing performance levels.

Despite these breakthroughs, the biological plausibility of modern DNNs remains in question. Fundamental practices, {\it e.g.}, non-local weight updates, forward-backward alignment, normalizers, advanced optimizers, and precise weight initialization, lack clear counterparts in the brain. While synaptic plasticity operates continuously with a ``batch size'' of one, DNNs typically rely on large batches and normalization layers~\cite{ioffe2015batch, ba2016layer} to stabilize activations. Optimizers like Adam~\cite{kingma2014adam} and RMSProp~\cite{hinton2012rmsprop} incorporate momentum-like processes and auxiliary state variables whose biological equivalents are speculative. Furthermore, although there is indirect evidence for synaptic scaling with local neural density~\cite{barral2016synaptic}, no biological mechanism matches the careful strategies used to initialize modern DNNs~\cite{glorot2010understanding,he2015delving}. \nst{These engineered solutions often increase memory consumption, which poses particular challenges for neuromorphic computing systems with limited resources.} Finally, deep networks often demand large datasets and struggle in the low data regime. This limitation starkly contrasts the remarkable data efficiency of biological learners~\cite{frank2023bridging}.

In DNNs, gradient-based methods efficiently solve the credit assignment problem, determining the weight's contribution to the network's error~\cite{richards2019deep}. Likewise, recent biologically plausible models have made strides in approximating these rules~\cite{lillicrap2016random}. Nevertheless, whether the brain relies on similar strategies or alternative mechanisms remains an open question~\cite{lillicrap2020backpropagation}. For instance, local plasticity rules such as Hebbian mechanisms~\cite{hebb2005organization} are believed to provide a biological basis for learning without explicit error signals. Still, Hebbian learning is prone to unrestrained weight growth. In this context, constrained Hebbian rules such as Oja's rule~\cite{oja1982simplified, oja1989neural} and related principal subspace methods~\cite{xu1993least, sanger1989optimal, karhunen1995generalizations} illustrate how plasticity rules stabilize learning without requiring an explicit normalization step. \nst{These rules converge to rotated bases in the eigenvector subspace of the input covariance matrix.} Nonetheless, learning goals may only be reached when the gradient is partially followed~\cite{richards2019deep}. Thus, Hebbian rules alone may fall short, as they ignore each synapse's impact on network's loss.

Here, we ask whether biological learning might be poised for a renaissance analogous to that of artificial networks a decade ago. We demonstrate that Oja's plasticity rule~\cite{oja1989neural}, combined with backpropagation, can reduce reliance on many algorithmic strategies with no clear biological analog. Oja's rule enables online learning in a manner that does not require batch normalization layers or precise weight initialization, allowing learning with fewer data points. \nst{By stabilizing network activations, this rule effectively replaces the benefits of standard normalization layers, even under the biologically realistic constraint of continuous (batch size $=1$) learning. Extending Oja's rule to recurrent neural networks reshapes network dynamics, stabilizing training and significantly enhancing short-term memory without biologically implausible gating. Furthermore, by improving layer-wise reversibility, Oja's rule helps reduce memory consumption, making it especially attractive for neuromorphic computing platforms.} 

Moreover, we find that Oja's rule synergizes with recently proposed weight-alignment approaches~\cite{akrout2019deep}, which use local plasticity rules to align forward and backward weights. When imperfectly aligned weights carry the global error, Oja's rule still enhances forward propagation, improving network performance under biologically plausible conditions. Remarkably, we show that meta-learning synaptic plasticity can uncover biologically motivated rules that, rather than merely \textit{matching} backpropagation's performance~\cite{payeur2021burst, shervani2023meta, akrout2019deep}, can \textit{exceed} it in low data regimes and where engineering ``tricks'' are absent or infeasible.

This work marks a step toward biologically inspired, high-performance learning algorithms. As such, it raises the possibility of another impending leap forward, this time in the realm of biologically plausible modeling.

\section*{Results}

We begin by considering a DNN with $L$ fully connected layers, defined by
\begin{equation}
    \mathbf{y}_\ell = \sigma(\mathbf{z}_\ell), 
    \quad
    \text{where}
    \quad
    \mathbf{z}_\ell = \boldsymbol{W}_{\ell-1,\ell}\,\mathbf{y}_{\ell-1}.
    \label{eq:fwd_pass}
\end{equation}
Here, $\boldsymbol{W}$ denotes the weight matrix, $\mathbf{y}$ the activation, $\mathbf{z}$ the pre-activation, $\sigma$ the nonlinear activation function, and $\ell$ indexes the layers. Let $\boldsymbol{X}_{\mathrm{train}} = \{\boldsymbol{x}^{(i)}_{\mathrm{train}}\}_{1 \leq i \leq N}$ represent the input training set and $\mathbf{y}_{\mathrm{train}} = \{y^{(i)}_{\mathrm{train}}\}_{1 \leq i \leq N}$ their corresponding labels. The loss function $\mathcal{L}(\mathbf{y}_L, \mathbf{y}_{\mathrm{train}})$ quantifies the discrepancy between the network's prediction $\mathbf{y}_L$ and ground-truth labels $\mathbf{y}_{\mathrm{train}}$.

Error backpropagation (backprop) optimizes $\boldsymbol{W}$ by propagating the gradient of $\mathcal{L}$ backward through each layer. Specifically, the gradient with respect to pre-activations, $\mathbf{e} = \partial \mathcal{L} / \partial \mathbf{z}$, flows recursively via
\begin{equation}
    \mathbf{e}_{\ell} = \boldsymbol{B}_{\ell+1,\ell} \,\mathbf{e}_{\ell+1} \,\odot\, \sigma'(\mathbf{z}_{\ell}),
    \label{eq:e_l_backprop}
\end{equation}
where $\odot$ denotes element-wise multiplication. In standard implementation of backprop, the matrix $\boldsymbol{B}_{\ell+1,\ell}$ is simply the transpose of the forward weight matrix, $\boldsymbol{W}_{\ell,\ell+1}^\top$. Once the error signals reach layer $\ell$, the parameters are updated as
\begin{equation}
    \Delta \boldsymbol{W}_{\ell-1,\ell} = -\theta \,\mathbf{e}_\ell \,\mathbf{y}_{\ell-1}^\top,
    \label{eq:BP}
\end{equation}
where $\theta$ is the learning rate.

Two standard modes of gradient-based training exist. In \emph{batch learning}, gradients are accumulated over mini-batches of data to reduce variance before a single update is applied, often yielding improved stability. By contrast, \emph{online learning} immediately updates parameters after each data sample arrives. The latter strategy is more biologically plausible, mirroring continuous synaptic plasticity in the brain. In this work, we focus on the online regime.

\subsection*{Oja's rule enhances online training of deeper networks}

\begin{figure}[H]
    \setlength{\unitlength}{0.01\textwidth}
    \begin{picture}(98,30)
        \put(2,0){\includegraphics[width=0.46\textwidth]{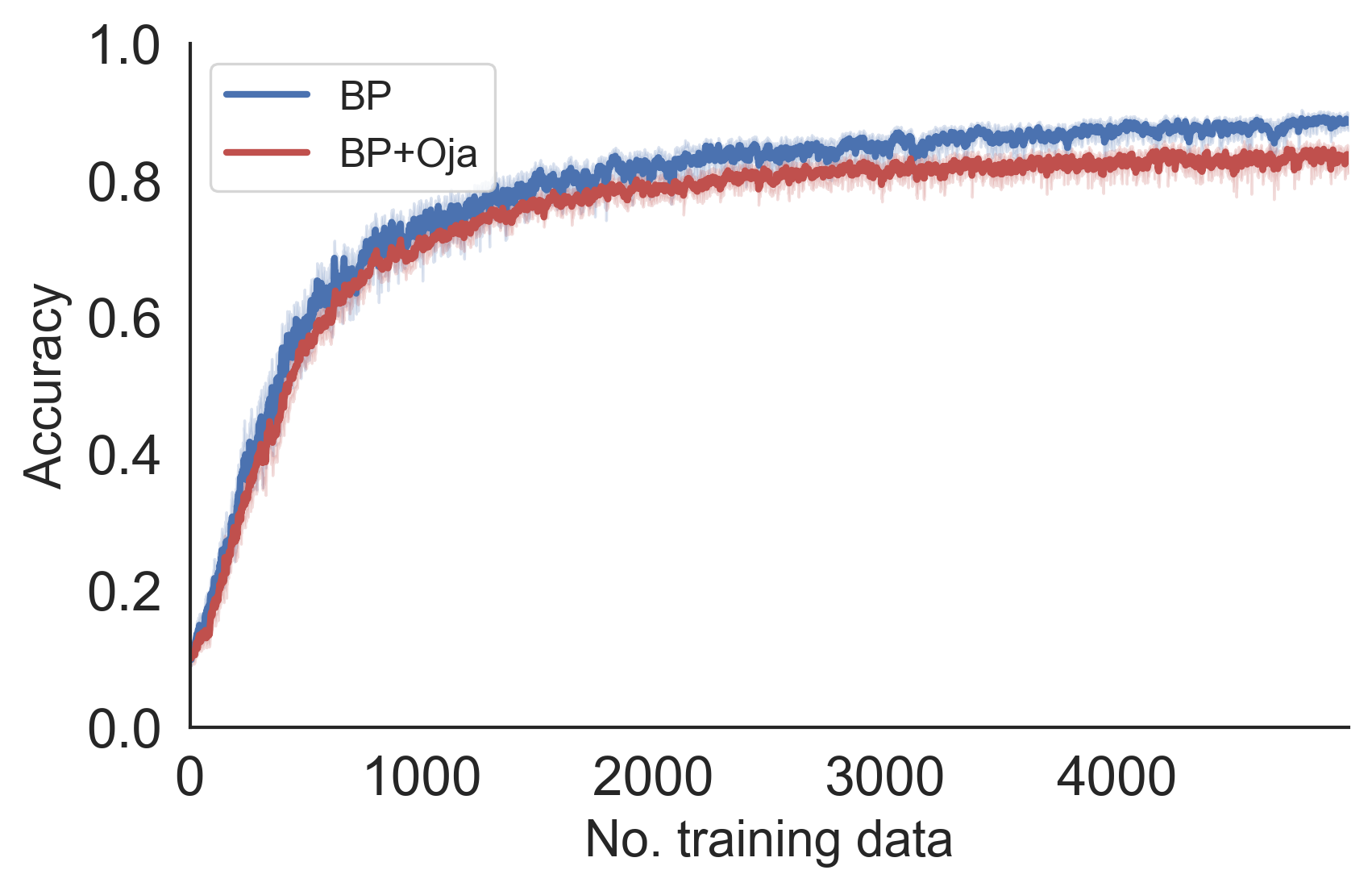}}
        \put(1,28){\footnotesize a}
        \put(23,28){5-layer}
        
        \put(52,0){\includegraphics[width=0.46\textwidth]{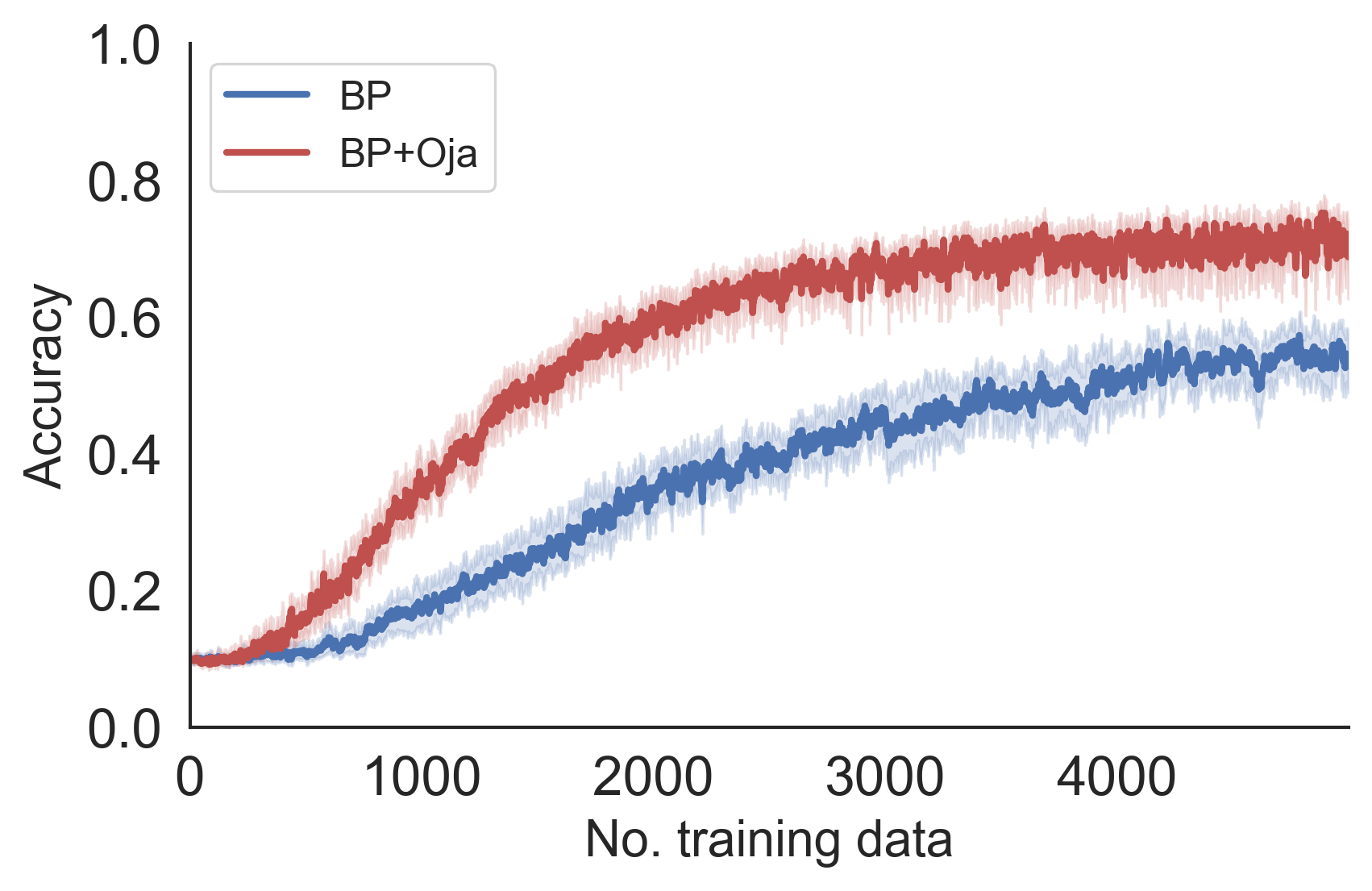}}
        \put(51,28){\footnotesize b}
        \put(73,28){10-layer}
    \end{picture}
    \caption{\textbf{Oja’s rule improves online learning in deep models.} \textbf{a}, MNIST validation accuracy during online training of a 5-layer network, and \textbf{b}, of a 10-layer network. BP (blue) and BP+Oja (red) curves represent averages over independent runs; shaded regions indicate $98\%$ confidence intervals. 
    }
    \label{fig:depth}
\end{figure}

We first consider a five-layer network trained online on the MNIST dataset, using a small subset of $N=5000$ samples. In this configuration, backprop alone achieves modest performance (Fig.~\ref{fig:depth}a, blue), which is below typical benchmarks due to both the smaller training set and the high variance introduced by online updates. As the network depth increases, backprop's performance drops noticeably (Fig.~\ref{fig:depth}b, blue), reflecting well-known difficulties in propagating error signals through many layers \cite{glorot2010understanding}. In the online setting, these challenges are further exacerbated by pronounced gradient variance.

Numerous engineering strategies, such as mini-batching, momentum terms \cite{qian1999momentum}, and batch normalization \cite{ioffe2015batch}, are effective at mitigating deep-learning instabilities. However, these methods generally lack the local, biologically plausible plasticity that is thought to underpin synaptic updates in the brain. In contrast, biological learning is often ascribed to Hebbian-like processes, which raise the question of whether Hebbian-inspired synaptic rules can stabilize online learning in deep architectures \cite{shervani2023meta}.

We therefore propose a hybrid update rule that combines local Hebbian plasticity with error-driven learning:
\begin{equation}
    \mathcal{F}^{\text{Oja}} 
    \;=\;
    \underbrace{-\theta_1 \,\mathbf{e}_{\ell} \,\mathbf{y}_{\ell-1}^\top}_{\text{backprop term}}
    \;+\;
    \underbrace{\theta_2 (\mathbf{y}_{\ell} \mathbf{y}_{\ell-1}^\top
    \;-\;
    (\mathbf{y}_{\ell} \mathbf{y}_{\ell}^\top)\,\boldsymbol{W}_{\ell-1,\ell})}_{\text{Oja's term}},
    \label{eq:BPOja}
\end{equation}
where $\theta_1$ and $\theta_2$ are learning rates chosen via grid search unless stated otherwise. The second term reflects Oja's rule \cite{oja1989neural}, providing Hebbian learning with built-in weight normalization.

For a five-layer network on MNIST, this hybrid rule achieves slightly lower accuracy than backprop alone (Fig.~\ref{fig:depth}a). Remarkably, for deeper networks (here, 10 layers), the hybrid rule provides a pronounced boost in validation accuracy compared to backprop alone (Fig.~\ref{fig:depth}b). Hence, the Oja-based term appears to mitigate challenges associated with online learning in deeper architectures, offering improved performance while retaining local, biologically plausible updates.

Oja's rule can be viewed as an online algorithm for principal subspace analysis (PSA)~\cite{xu1993least}. In its simplest form, PSA seeks an orthonormal set of basis vectors that minimize the residual between the original data and its projection onto a lower-dimensional subspace. A nonlinear extension of this measure may be written as
\begin{equation}
    \mathcal{J}_{\ell-1,\ell}
    \;=\;
    \frac{1}{N}\sum_{i=1}^N
    \bigl\|
    \mathbf{y}_{\ell-1}^{(i)}
    \;-\;
    \boldsymbol{W}_{\ell-1,\ell}^\top 
    \,\sigma\bigl(\boldsymbol{W}_{\ell-1,\ell}\,\mathbf{y}_{\ell-1}^{(i)}\bigr)
    \bigr\|^2.
    \label{eq:res_err}
\end{equation}
Crucially, Oja's rule emerges as a stochastic gradient descent on this residual~\cite{karhunen1994representation, xu1993least}.

\begin{figure}[H]
    \setlength{\unitlength}{0.01\textwidth}
    \begin{picture}(100,51)
      \put(1,49){\footnotesize a}
      \put(2,19){\includegraphics[width=0.47\textwidth]{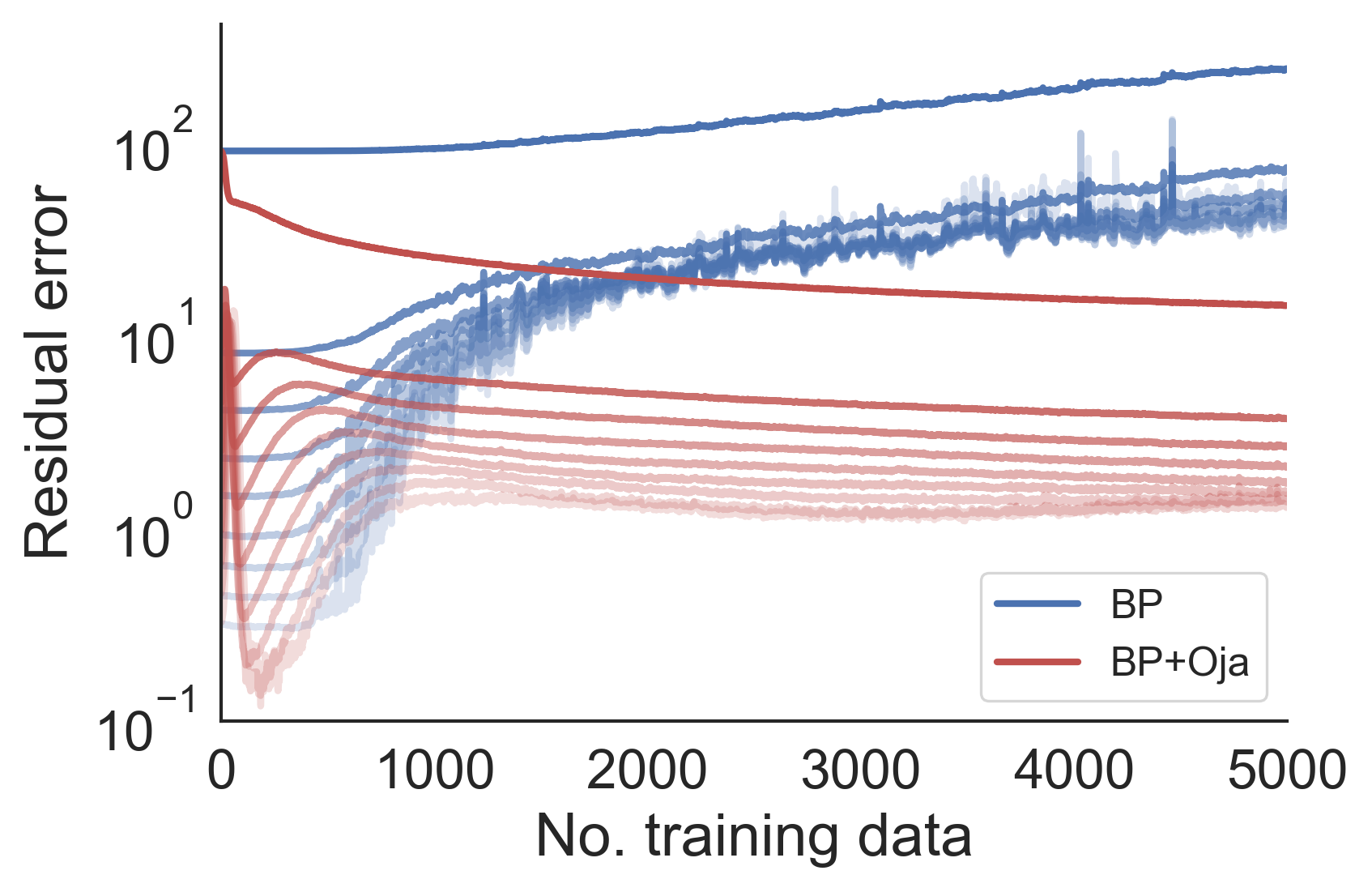}}
      \put(51,49){\footnotesize d}
      \put(52,19){\includegraphics[width=0.47\textwidth]{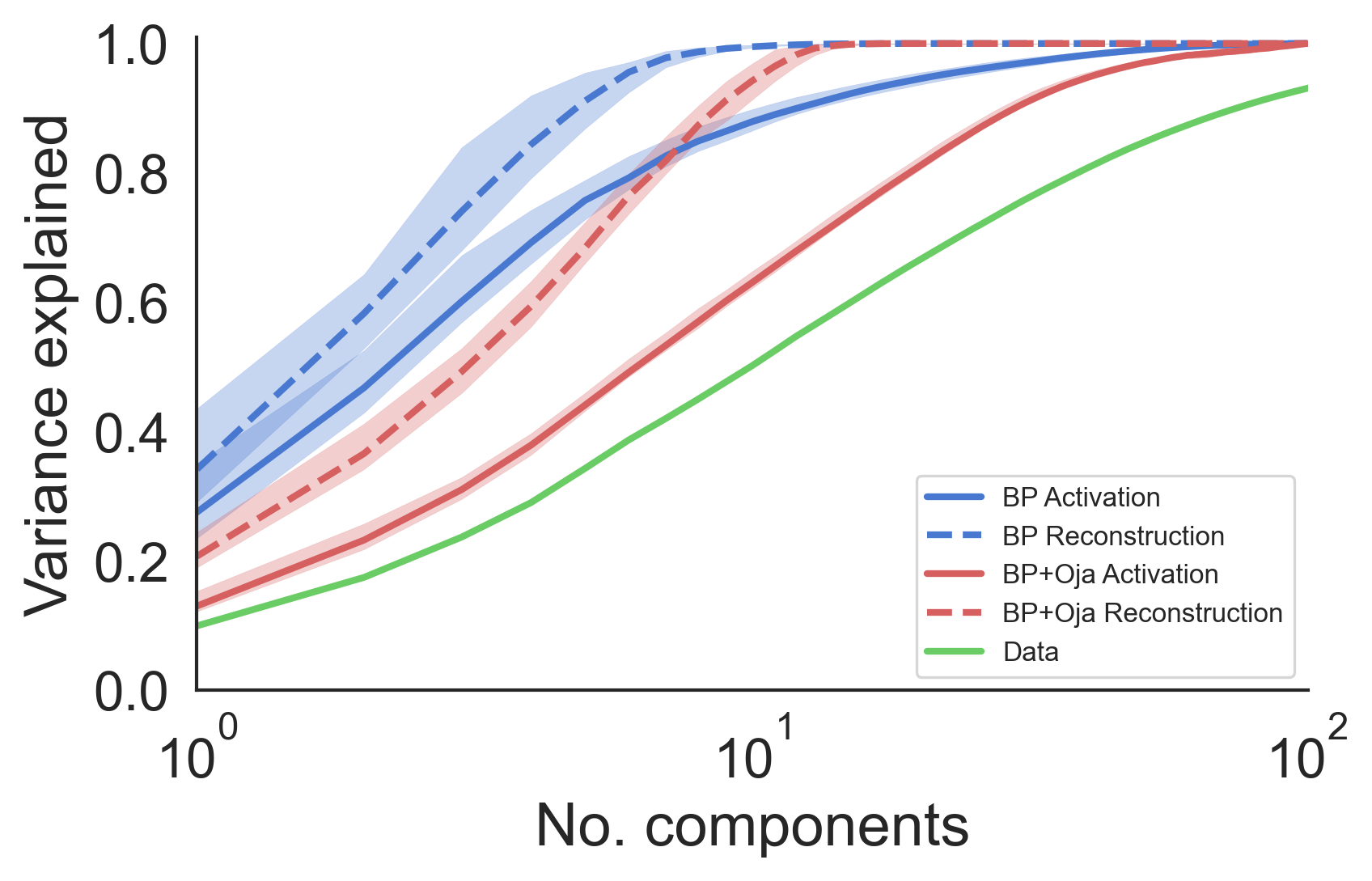}}
      \put(1,16.5){\footnotesize b}
      \put(2,4.5){\includegraphics[width=0.42\textwidth]{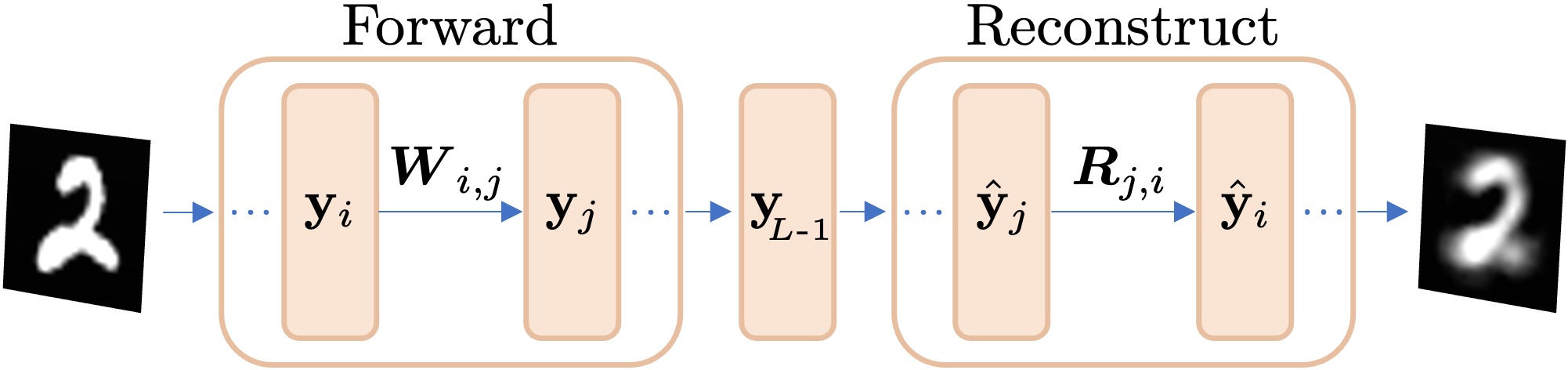}}
      \put(46,16.5){\footnotesize c}
      \put(47,0.5){\includegraphics[width=0.52\textwidth]{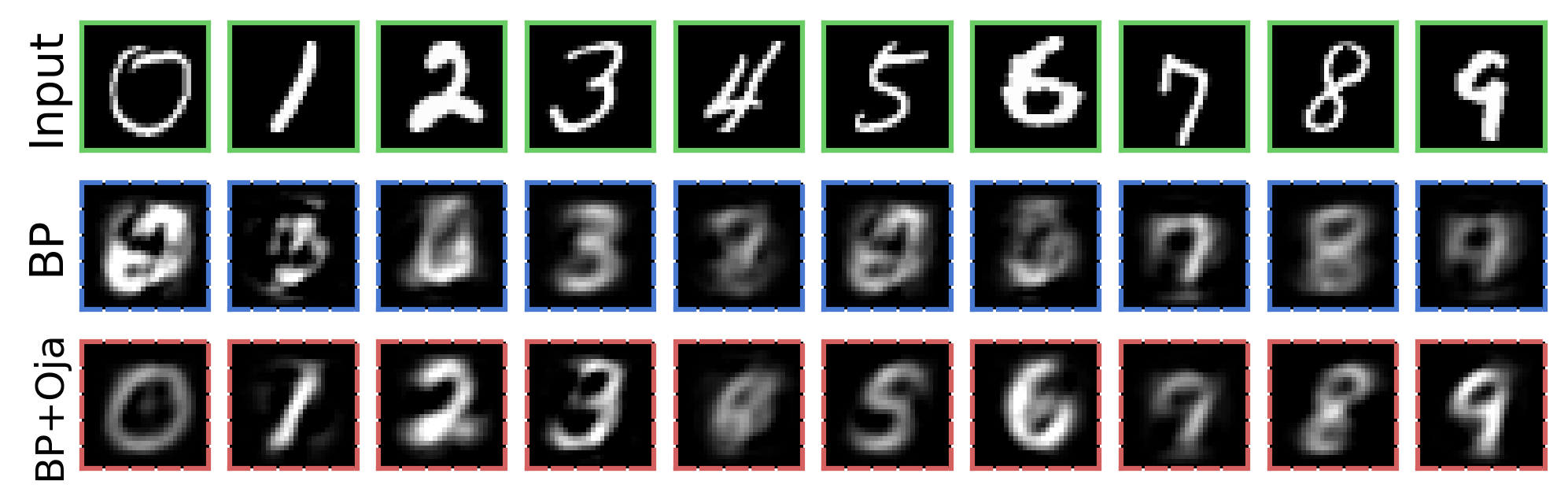}}
    \end{picture}
    \caption{\textbf{Oja's rule preserves richer information across deep layers.} \textbf{a}, Residual error (equation~(\ref{eq:res_err})) across layers of a 10-layer network trained using standard backprop (BP; blue) and $\mathcal{F}^{\text{Oja}}$ (BP+Oja; red). Decreasing brightness indicates deeper layers. \textbf{b}, Layer-wise reconstruction process. The input digit (left) is mapped forward through a cascade of nonlinear layers parameterized by $\boldsymbol{W}_{i, j}$ (Forward) culminating in the final hidden representation $\mathbf{y}_{L-1}$. To evaluate how well $\mathbf{y}_{L-1}$ retains information from the original input, we use optimal linear maps $\boldsymbol{R}_{j, i}$ to recursively reconstruct upstream activations $\hat{\mathbf{y}}_i$ (Reconstruct) and ultimately recover the input (right).
    \textbf{c}, Input images (top row; green boarder) and their reconstructions for a 10-layer network trained via backprop-only (middle row; blue) and combined backprop and Oja (bottom row; red). The BP+Oja model retains more character-defining information (e.g., “2” is not confused with “6”). 
    \textbf{d}, Cumulative explained variance for three sets of data: the original input (green), first-layer activations $\mathbf{y}_1$ (solid lines), and the reconstructed first-layer activations $\hat{\mathbf{y}}_1$ obtained from the final layer (dashed lines). Updating via $\mathcal{F}^{\text{Oja}}$ spreads variance more evenly across principal components and stays closer to the input curve, suggesting a richer, higher-dimensional subspace. 
    }
    \label{fig:residual}
\end{figure}

When transformations at intermediate layers become too lossy, the propagated signals can degrade beyond usefulness for downstream classification. Thus, a critical question is whether $\mathcal{F}^{\text{Oja}}$ can retain the optimal projection benefits of Oja's rule, while performing error-based learning. We therefore evaluated the residual error (equation~(\ref{eq:res_err})) across all layers in a 10-layer network trained either by backprop or with $\mathcal{F}^{\text{Oja}}$. Incorporating the Oja's term substantially reduced residual error in each layer (Fig.~\ref{fig:residual}a), indicating improved fidelity in deep representations.

To measure how well post-synaptic activations at each layer preserve upstream information, we fit an optimal linear map $\boldsymbol{R}_{\ell,\ell-1}$ that restores activations in layer $\ell-1$ from those in layer $\ell$ (Fig.~\ref{fig:residual}b; see Methods). The reconstructed images from the final hidden layer of two 10-layer networks are shown in Fig.~\ref{fig:residual}c. Under backprop, severe distortions often occurred (e.g., a digit “2” morphing into a “6”), implying that higher layers collapsed distinct inputs onto similar representations. By contrast, the hybrid update retained substantially clearer digit identities.

Further insights come from examining the cumulative explained variance under princiapl component analysis (PCA) for early-layer activations and their reconstructions. The hybrid model spread variance more evenly across principal components and remained closer to the intrinsic variability of the input data (Fig.~\ref{fig:residual}d; see Supplementary note \ref{sec:ve}). Combined, results in Fig.~\ref{fig:residual} suggest that Oja's update helps maintain a richer, higher-dimensional latent representation of inputs. In deeper architectures, this more faithful preservation of input structure ultimately improves both stability and classification performance.

\subsection*{Oja's rule avoids the need for precise weight initialization}

DNNs are notoriously sensitive to their initial weight configurations, motivating a wealth of research on principled initialization procedures. Common strategies, such as Xavier~\cite{glorot2010understanding} or Kaiming initialization~\cite{he2015delving}, ensure that activations and gradients remain well-scaled by drawing weights from precisely defined distributions dependent on layer dimensionality. These methods have become the de facto standard in deep learning libraries (e.g., PyTorch~\cite{paszke2019pytorch}, TensorFlow~\cite{abadi2016tensorflow}), effectively “baking in” the assumption that careful initialization is essential and readily accessible. Yet, this assumption lacks direct biological plausibility, prompting whether such precise initialization is strictly necessary for successful learning in neural networks.

An alternative class of initialization methods employs an iterative pre-processing stage. These methods adapt weights through a dedicated pre-training phase before gradient-based optimization. Notably, Mishkin and Matas~\cite{mishkin2015all} demonstrated that a refined initialization alone could improve learning outcomes, pithily summarized as ``All you need is a good init.'' Earlier, Karayiannis~\cite{karayiannis1996accelerating} showcased the feasibility of two-stage protocols by applying Oja's rule in an unsupervised pre-training step, followed by standard backprop. Departing from such assumptions on initial weights, we posit that a suitably chosen plasticity rule obviates the need for any dedicated initialization. In other words, ``all you need is a good plasticity rule.''

\begin{figure}[H]
    \centering
    \includegraphics[width=.49\textwidth]{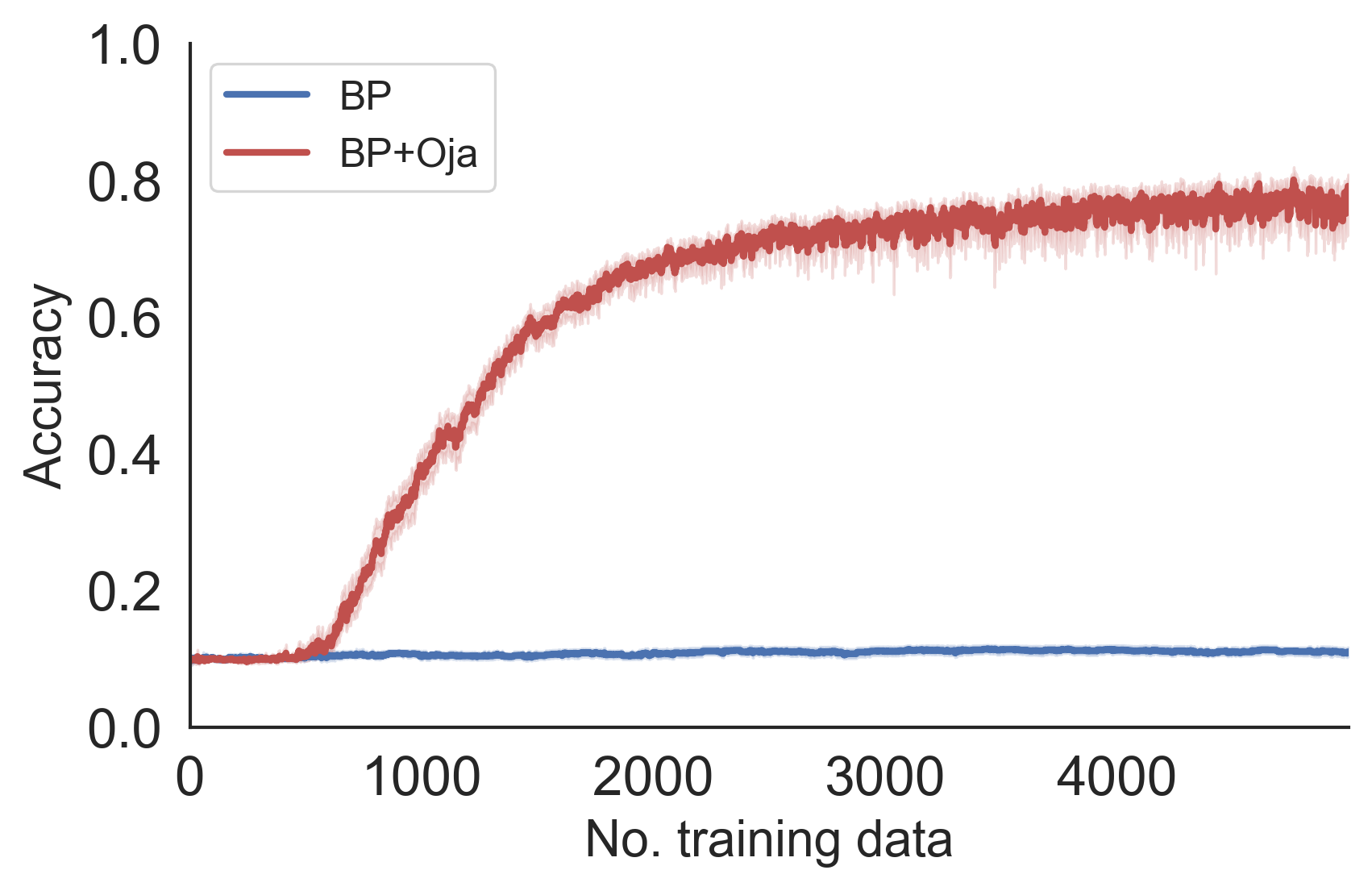}
    \caption{\textbf{Oja's rule improves learning in poorly initialized networks.} Validation accuracy on MNIST classification in a five-layer network with naive weight initialization, trained using backprop alone (BP, blue) and backprop combined with Oja's rule (BP+Oja, red). }
    \label{fig:naive}
\end{figure}

To test this idea, we trained a five-layer DNN (identical in architecture to Fig.~\ref{fig:depth}a) using a simplistic ``naive'' initialization scheme~\cite{krizhevsky2012imagenet}, where weights are drawn from the normal distribution $\mathcal{N}(0, 0.01)$. This approach doesn't regulate activation magnitudes, making it less refined than modern protocols. As expected, relying on backprop alone yielded minimal improvement in the accuracy of the classifier network, even after 5,000 training samples (Fig.~\ref{fig:naive}, blue curve). Poorly scaled weights generate weak gradients and diminishing signals that compound through successive layers, effectively stalling learning processes~\cite{lecun2012efficient, he2015delving} (see Supplementary note~\ref{sec:supp_oja_vs_bp}).

Incorporating Oja's rule alongside backprop dramatically enhanced performance (Fig.~\ref{fig:naive}, red curve). Oja's rule continually adjusts weights according to the evolving distribution of neural activations, compensating for suboptimal initial conditions in a biologically tenable manner. By dynamically regulating the norm of the synaptic weights during training, Oja's rule preserves signal flow throughout the network and promotes stable gradient propagation, even in deeper architectures. Thus, what initially appears to be a need for meticulously engineered weight initialization may be overcome by an adaptive plasticity mechanism that automatically normalizes connections online. In doing so, Oja's rule offers a compelling alternative to stage-based biologically motivated initialization routines and precisely calibrated initial weight distributions.

\subsection*{Oja's rule performs normalization online}

A key strategy for ensuring stable activations during neural network training is incorporating normalization techniques that directly adjust activations within the model's architecture. Batch normalization (BN)~\cite{ioffe2015batch} is a widely used example, normalizing mini-batch statistics (mean and variance) at each layer to maintain robust feature distributions. Though practical, this process introduces additional parameters (mean and variance for each output dimension of linear layers) and demands memory for storing these statistics, which is biologically complex. It also relies on batch-based operations that are inconsistent with purely online learning scenarios.   

By contrast, Oja's rule indirectly controls activations by modifying weight vectors' norms and orientation during training. After each weight update, the forward pass has no separate or explicit normalization step. Instead, the orthonormality constraint imposed by Oja's rule stabilizes activations and prevents them from growing excessively large. We provide formal proof in the Supplementary note~\ref{sec:norm_proof} showing that, for linear networks, the minimization of the residual error (equation~(\ref{eq:res_err})) by Oja's rule enforces preserving the activations' Euclidean norm across layers (i.e., $\|\mathbf{y}_{\ell}\| = \|\mathbf{y}_{\ell-1}\|$). While extending this derivation to nonlinear architectures poses analytical challenges, our empirical findings confirm that Oja's rule similarly preserves activation norms in deep networks with nonlinearities (Fig.~\ref{fig:norm}\nst{; see Supplementary Fig.~\ref{fig:sup_y_norms}}). Intuitively, BN explicitly stabilizes the mean and variance (and thus indirectly affects activation norms), whereas Oja's rule enforces direct norm preservation, thereby stabilizing both mean and variance.

\begin{figure}[H]
    \centering
    \includegraphics[width=0.49\textwidth]{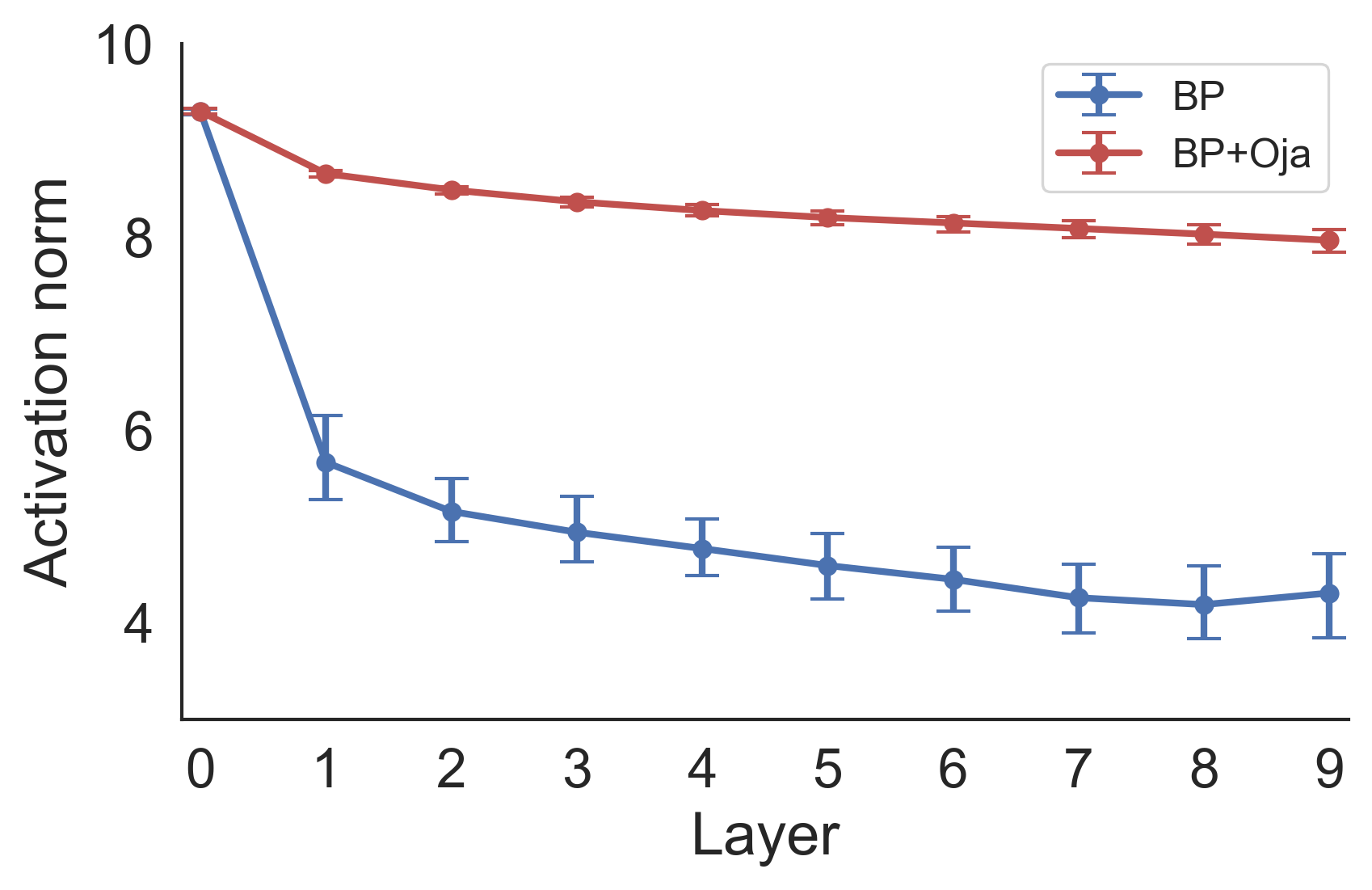}
    \caption{\textbf{Oja's rule preserves activation norms across layers.} Activation norms (y-axis) are shown across the nine hidden layers and input (x-axis) of a neural network trained with standard backprop (BP, blue) and with backprop supplemented by Oja's rule (BP+Oja, red). Each point represents the average activation norm over the test set, computed after 5,000 online training iterations. Error bars indicate $98\%$ confidence interval. Introducing Oja's rule maintains the activation norm closer to the input layer, while BP alone shows a notable decrease in norm and more variability across runs.}
    \label{fig:norm}
\end{figure}

\subsection*{Oja’s rule reshapes the eigenspectrum and improves short-term memory in RNNs}

Neural circuits in the brain are intrinsically recurrent, forming feedback loops that enable the integration of information over time. In machine learning, recurrent neural networks (RNNs) harness similar feedback mechanisms to process sequential data but often face challenges in retaining information across extended sequences \cite{bengio1994learning,pascanu2013difficulty}. Strategies to mitigate these limitations range from introducing gating mechanisms and memory cells \cite{hochreiter1997long} to careful weight initialization with unitary or orthogonal structures \cite{le2015simple,henaff2016recurrent}. While effective, these methods frequently depart from biologically plausible principles \cite{costa2017cortical}.

We propose an alternative approach that augments standard backprop through time (BPTT) with Oja's rule. To incorporate biologically inspired weight normalization, we embed Oja's rule into the recurrent connections' update. We let $\mathbf{y}_t = \boldsymbol{W}_h\,\sigma(\mathbf{h}_{t-1})$ denote the forward temporal mapping to the hidden units. Here, $\sigma$ denotes the nonlinear activation function and $\boldsymbol{W}_h$ is the recurrent weight matrix. The combined weight update is then given by
\begin{equation}
\mathcal{F}\,^{\text{Oja}}_h 
\;=\; 
\underbrace{\theta_1 \sum_{t=1}^T\mathbf{e}_{t}\,\sigma(\mathbf{h}_{t-1})^\top}_{\text{BPTT term}} 
\;+\; 
\underbrace{\theta_2 \sum_{t=1}^T\Bigl(\mathbf{y}_t\,\mathbf{h}_{t-1}^\top 
\;-\; 
\bigl(\mathbf{y}_t\,\mathbf{y}_t^\top\bigr)\,\boldsymbol{W}_h \Bigr)}_{\text{Oja's term}},
\end{equation}
where $\mathbf{e}_{t}$ represents the backpropagated error signal at time step $t$, and $\theta_1$ and $\theta_2$ are the learning rates (see Methods for recurrent dynamics).

\begin{figure}[H]
  \setlength{\unitlength}{0.01\textwidth}
  \begin{picture}(98,81)
    \put(0,80){\footnotesize a}
    \put(2,61){\includegraphics[width=0.49\textwidth]{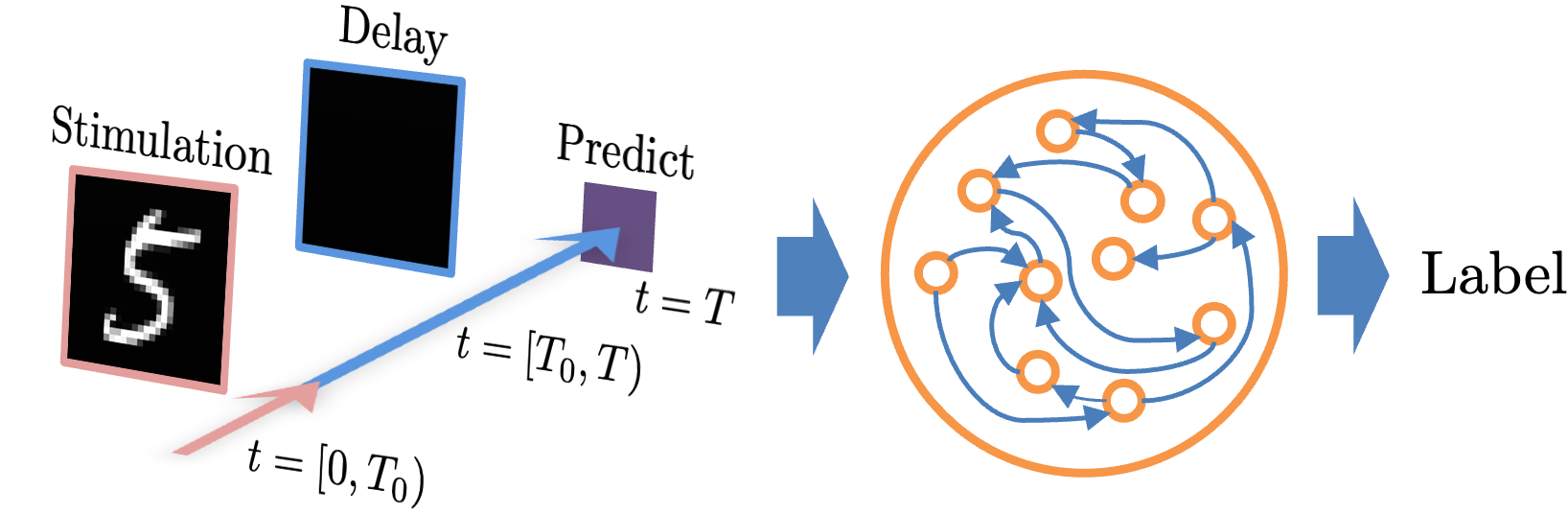}}

    \put(55.5,80){\footnotesize b}
    \put(56.5,55){\includegraphics[width=0.40\textwidth]{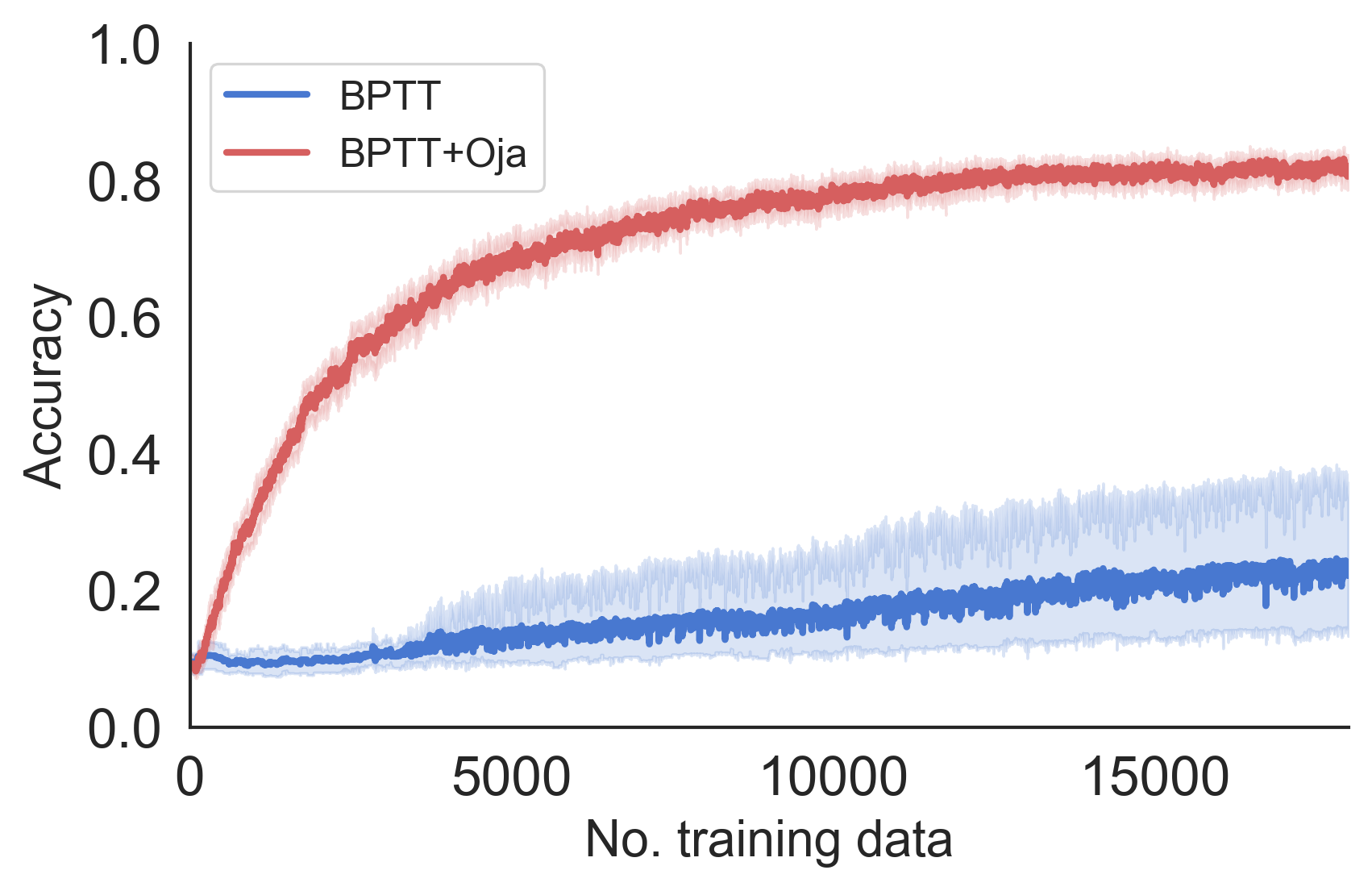}}
    
    \put(0,52.5){\footnotesize c}
    \put(1,23){\includegraphics[width=0.47\textwidth]{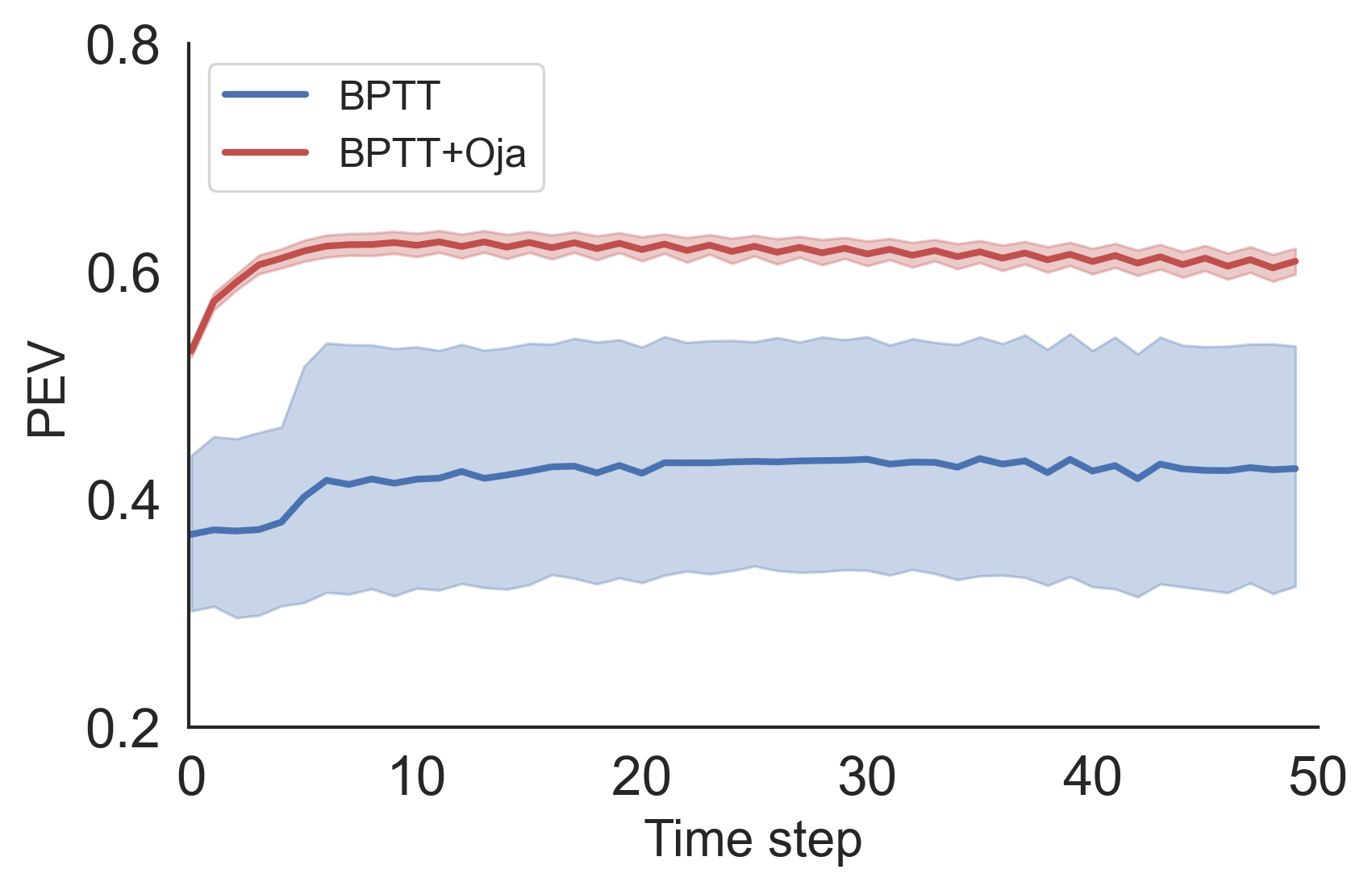}}
    
    \put(49.5,52.5){\footnotesize d}
    \put(50.5,23){\includegraphics[width=0.47\textwidth]{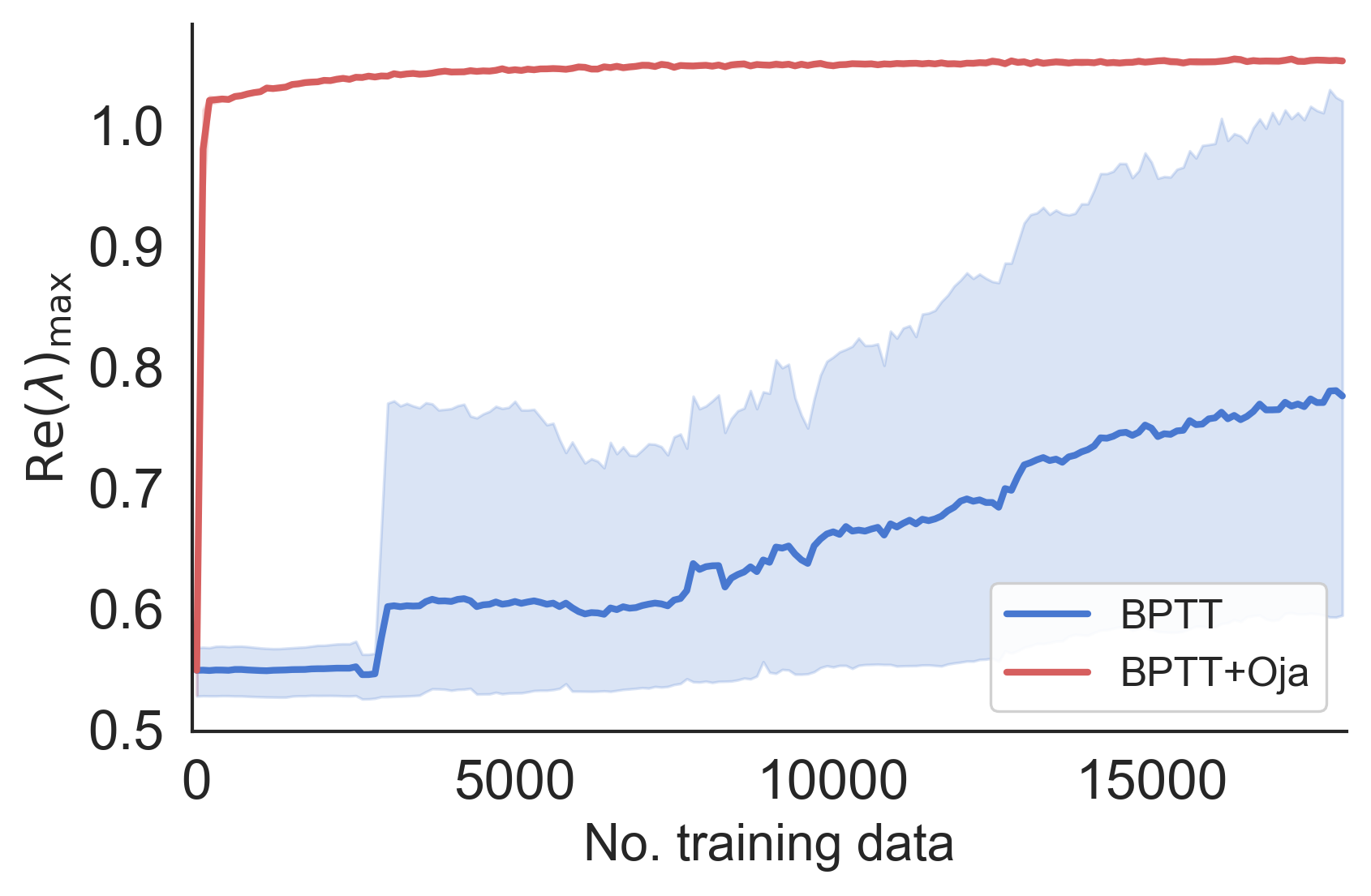}}
    
    \put(0,21){\footnotesize e}
    \put(2,0){\includegraphics[width=0.97\textwidth]{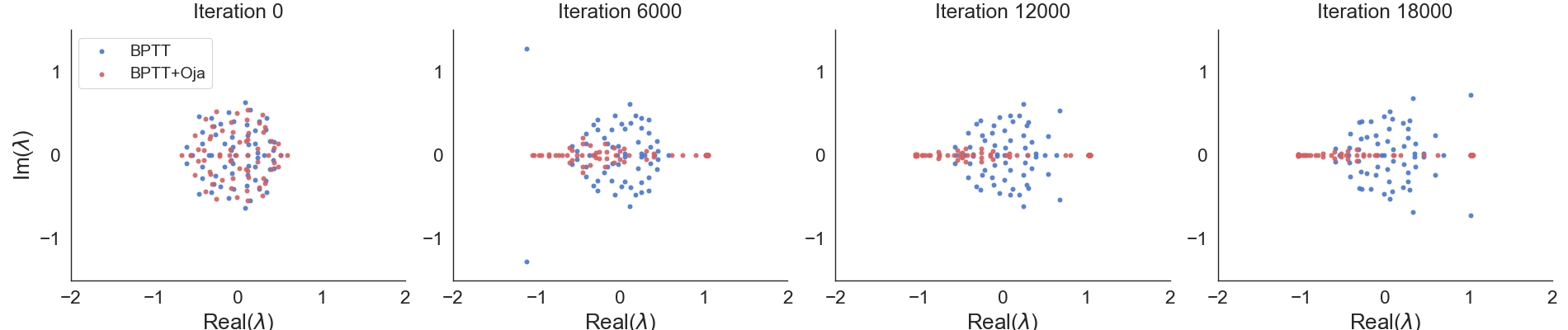}}
  \end{picture}
  \caption{\textbf{Oja's rule improves short-term memory and stabilizes RNN dynamics.} Validation performance and dynamical stability of an RNN trained on a delayed MNIST classification task are compared for standard Backprop Through Time (BPTT) and BPTT combined with Oja's rule (BPTT+Oja). \textbf{a}, Task schematic: each trial begins with an MNIST digit presentation, followed by a delay, after which the model predicts the label. 
  \textbf{b}, Classification accuracy over training iterations.
  \textbf{c}, Proportion of explained variance (PEV) over time, indicating the fraction of hidden-state variance attributable to digit identity.
  \textbf{d}, Evolution of the maximum real eigenvalue, $\operatorname{Re}(\lambda)_{\max}$. 
  \textbf{e}, Superimposed eigenvalue distributions of the recurrent weight matrix at four training snapshots for a representative run (highest‑performing seed across independent runs for the baseline BPTT, randomly selected seed for BPTT+Oja); under BPTT+Oja the eigenvalues collapse onto the real axis and the spectral radius is approximately bounded by 1.}  
  \label{fig:RNN}
\end{figure}

We evaluated this hybrid training scheme on a delayed classification task (Fig.~\ref{fig:RNN}a). In each trial, an MNIST image is presented to the network for $T_0=5$ time steps, followed by a delay until $T=50$, at which point the network classifies the digit via the final readout. Notably, even with modest hidden-state dimensions, RNNs trained with BPTT augmented by Oja's rule (BPTT+Oja) achieve significantly higher accuracy than those trained using BPTT alone (Fig.~\ref{fig:RNN}b). Moreover, analysis of the proportion of explained variance (PEV) reveals that the BPTT+Oja model not only encodes higher stimulus-related information content but also preserves it more stably over extended delays (Fig.~\ref{fig:RNN}c; see Methods).

Beyond accuracy gains, incorporating Oja's rule profoundly reshapes the dynamical properties of the recurrent weight matrix. We track the maximum real eigenvalue, $\mathrm{Re}(\lambda)_{\max}$, as training progresses. Under the hybrid regime, $\mathrm{Re}(\lambda)_{\max}$ converges near unity with low variance (Fig.~\ref{fig:RNN}d), whereas the BPTT-only model exhibits substantial variability. Our results further demonstrate that, under $\mathcal{F}\,^{\text{Oja}}_h$, the imaginary components of the eigenvalues are systematically driven toward zero (Fig.~\ref{fig:RNN}e). In contrast, the eigenvalues in the BPTT model retain substantial imaginary components throughout training. This “collapse” of eigenvalues onto the real axis parallels findings in biologically plausible models~\cite{sheeran2024spatial}. In sum, the integration of Oja's rule steers the network toward alternative dynamical regimes that improve short-term memory performance, obviating the need for highly engineered architectures or specialized initialization schemes.

\subsection*{Oja’s rule improves network performance without explicit weight alignment}

In the previous sections, we showed that removing common engineering shortcuts from backprop degrades learning and that Oja's rule alone can replace many of these heuristics. Although backprop remains biologically implausible (e.g., requiring symmetric forward and backward weights; $\boldsymbol{B}_{\ell+1,\ell}=\boldsymbol{W}_{\ell,\ell+1}^\top$)~\cite{lillicrap2020backpropagation}, this controlled setup confines learning inefficiencies to the lack of stabilization rather than the plasticity mechanism itself. Thus, Oja's rule provides broadly applicable stabilizing dynamic that remains valuable alongside an already \nst{efficacious} training procedure. We now turn to scenarios without explicit weight symmetry.

\begin{figure}[H]
    \setlength{\unitlength}{0.01\textwidth}
    \begin{picture}(98,62)
    \put(0,61){\footnotesize a}
    \put(1,31){\includegraphics[width=0.47\textwidth]{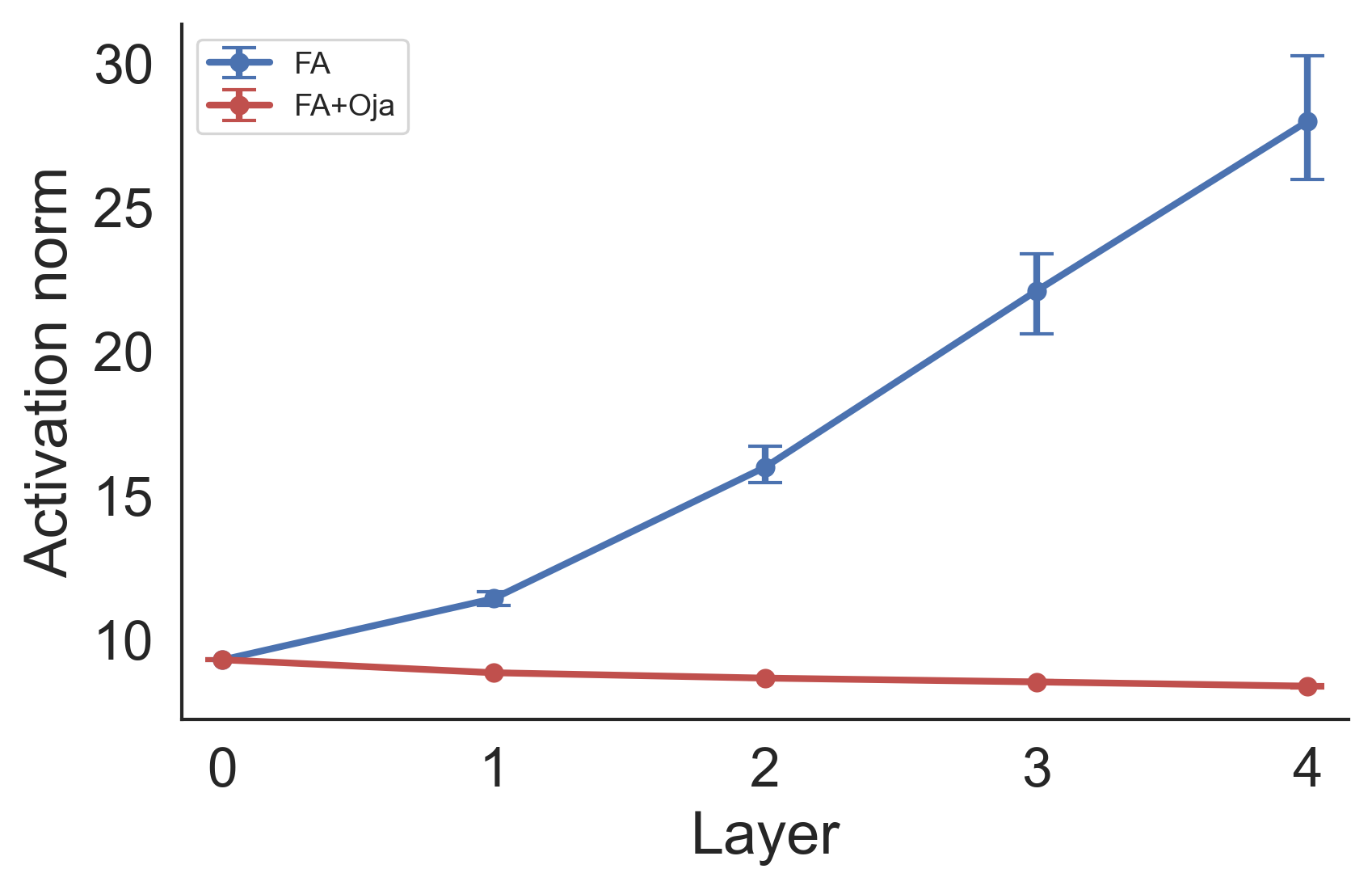}}

    \put(50,61){\footnotesize b}
    \put(57,36){\includegraphics[width=0.39\textwidth]{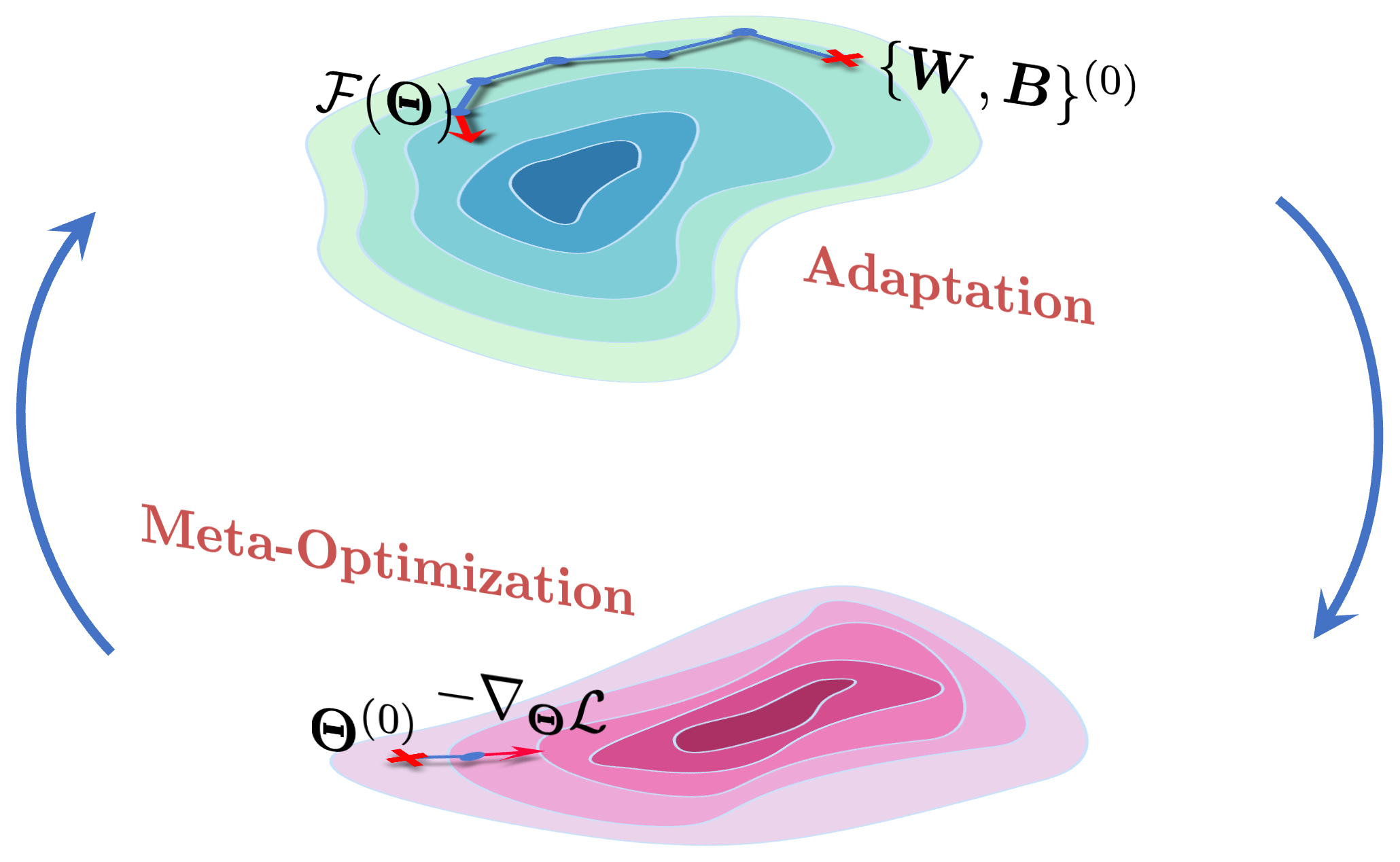}}
    
    \put(0,29){\footnotesize c}
    \put(1,0){\includegraphics[width=0.47\textwidth]{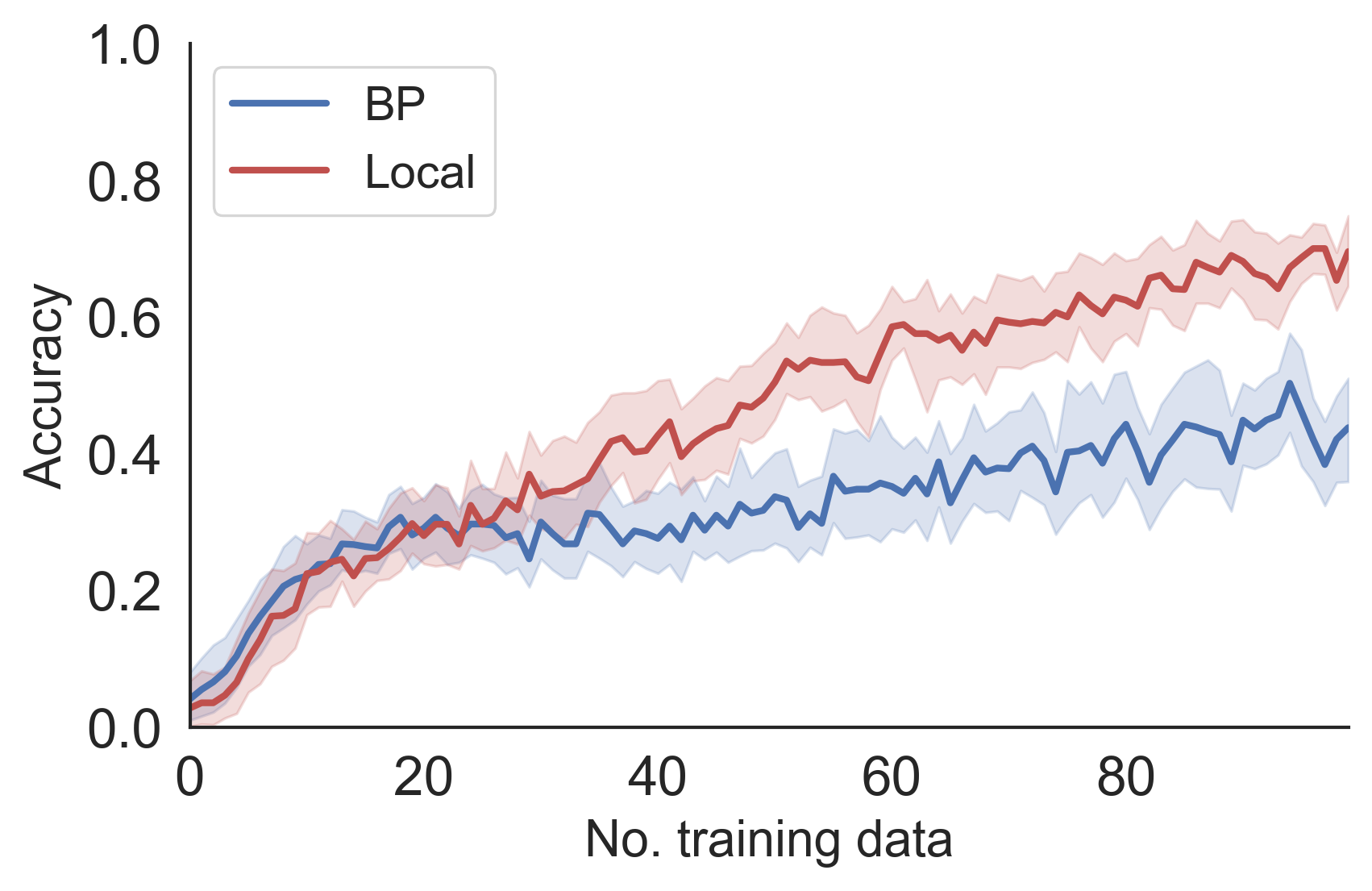}}
    
    \put(49.5,29){\footnotesize d}
    \put(50.5,0){\includegraphics[width=0.47\textwidth]{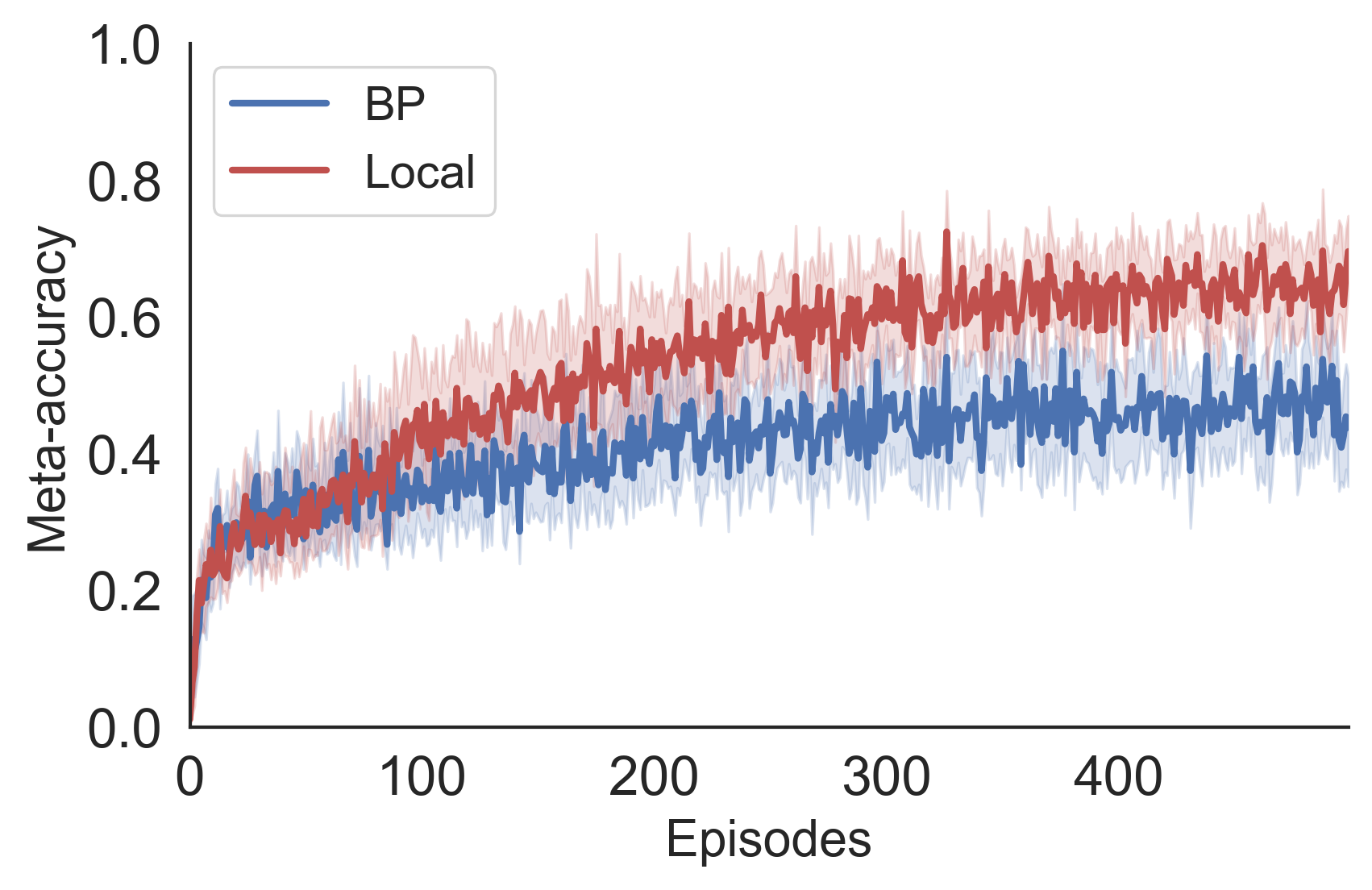}}
  \end{picture}
    \caption{\textbf{Oja's rule gives rise to biologically plausible rules that can outperform backprop.}
    \textbf{a}, Activation norms across layers after training under feedback‐alignment (FA, blue) versus FA combined with Oja (red). Oja's rule prevents runaway growth in activations.
    \textbf{b}, Schematic of the nested meta‐learning procedure: An inner adaptation loop (top) trains a randomly initialized network on 100 samples (5‐class classification) via a parameterized rule $\mathcal{F}(\boldsymbol{\Theta})$, while an outer meta‐optimization loop (bottom) updates the rule's coefficients $\boldsymbol{\Theta}$ by differentiating through the unfolded adaptation computational graph.
    \textbf{c}, Adaptation accuracy at the final episode, showing how the converged plasticity rule trains the network from random initial conditions.
    \textbf{d}, Meta‐accuracy across episodes, comparing the performance of the local rule (red; equation~(\ref{eq:delta_B})) and backprop (blue; equation~(\ref{eq:BP})) as the plasticity parameters are refined. Each episode represents a full cycle of inner adaptation followed by a meta-parameter update.
    }
    \label{fig:BioOja}
\end{figure}

Meta‑learning has emerged as a robust framework for discovering effective, interpretable learning rules under biological constraints~\cite{confavreux2020meta, shervani2023meta, bell2024discovering}. In earlier work~\cite{shervani2023meta}, we applied this approach to identify plasticity rules that improve performance under fixed random feedback alignment (FA)~\cite{lillicrap2016random}, showing that incorporating Oja's rule offsets the limitations of fixed backward pathways. Here, we demonstrate that Oja's primary contribution to FA is stabilizing activation norms rather than enforcing strict orthogonality, which becomes infeasible with nonlinearities and error signals. By preventing unbounded norm growth, Oja's rule averts exponential blowups in downstream layers caused by cumulative upstream expansion under FA (Fig.~\ref{fig:BioOja}a).

Next, we remove the constraint of fixed feedback and aim to outperform backprop without relying on biologically dubious weight symmetry. Using a nested meta-learning algorithm (Fig.~\ref{fig:BioOja}b; see Methods), we jointly optimize local updates in both forward and backward pathways. In the inner loop, randomly initialized networks adapt to small training sets via a parameterized rule, $\mathcal{F}(\boldsymbol{\Theta})$, where $\boldsymbol{\Theta}$ denotes meta-parameters. An outer meta-optimizer then refines $\boldsymbol{\Theta}$ by differentiating through the entire unrolled adaptation pipeline. Formally, $\mathcal{F}(\boldsymbol{\Theta}) = \{\Delta \boldsymbol{W},\, \Delta \boldsymbol{B}\}$, with 
\begin{align}
\begin{split}
\Delta \boldsymbol{W}_{\ell-1,\ell}
=-&\theta_0 \mathbf{e}_{\ell}\mathbf{y}_{\ell-1}^\top + \theta_1(\boldsymbol{1}_{\ell} \mathbf{y}_{\ell-1}^\top - \mathbf{y}_{\ell} \mathbf{1}_{\ell-1}^\top) + \theta_2(\mathbf{y}_{\ell} \mathbf{e}_{\ell-1}^\top - \mathbf{e}_{\ell} \mathbf{1}_{\ell-1}^\top) \\
+& \theta_3 (\mathbf{y}_{\ell}\mathbf{y}_{\ell-1}^\top - (\mathbf{y}_{\ell}\mathbf{y}_{\ell}^\top) \boldsymbol{W}_{\ell-1,\ell}) -\theta_4 \boldsymbol{W}_{\ell-1,\ell}\\
\Delta \boldsymbol{B}_{\ell,\ell-1} = -&\theta_5 \mathbf{y}_{\ell}\mathbf{e}_{\ell-1}^\top+\theta_6 \boldsymbol{B}_{\ell,\ell-1},
\end{split}
\label{eq:delta_B}
\end{align}
and $\boldsymbol{\Theta}=\{\theta_r\,|\,0\leq r \leq 6\}$ representing the plasticity meta‑parameters. Notably, the learned rules naturally incorporate Oja's rule to stabilize neural activations. As shown in Figs.~\ref{fig:BioOja}c--d, both the final adaptation accuracy (panel c) and meta‑accuracy over repeated episodes (panel d) exceed those achieved by standard backprop.

These findings deviate from the prevailing efforts that design plasticity rules to align weights as the primary strategy for efficient credit assignment in deep networks~\cite{shervani2023meta,akrout2019deep}. Instead, Oja's rule improves learning by enhancing the forward information flow and maintaining rich, stable representations rather than enforcing precise alignment between forward and backward weights. \nst{While the previous works have primarily aimed at matching the performance of backprop,} this result highlights that under biologically realistic constraints (online updates, small data regimes, and non-symmetric feedback), local plasticity rules can outperform backprop.

\newpage
\section*{Discussion}

Despite the undeniable success of error-driven training, the biological plausibility of modern deep neural networks remains a pressing concern. In essence, the engineering “tricks” that empower DNNs have few direct analogs in the brain. Our study provides evidence that Oja’s plasticity rule robustly addresses key barriers typically encountered when training neural networks under biologically plausible conditions. Previous work established that incorporating Oja’s rule could elevate biologically inspired error-driven learning to match classical backprop~\cite{shervani2023meta}. Here, we extend these findings by demonstrating not only parity but also superiority over pure backprop in challenging scenarios such as deeper architectures, naive weight initialization, the absence of normalization techniques, and online, biologically realistic learning environments.

Earlier research primarily examined Oja’s rule in isolation and within strictly unsupervised contexts. In contrast, we explicitly integrate it with backprop and error-driven schemes. This pairing creates an online regularizer that adaptively adjusts synaptic weights in proportion to neural activity. The hybrid rule effectively counteracts destabilizing factors during training, preserves the integrity of signal propagation throughout the model, and maintains the fidelity of representations across layers.

Oja’s rule inherently enforces orthonormality, aligning the hybrid rule with manifold-based learning schemes that uphold these constraints. Existing methods for learning under orthonormal constraints include projected gradient techniques that map updates on the Stiefel manifold via costly retraction steps~\cite{edelman1998geometry} and gradual methods that guide weight configurations toward a near-orthonormal regime without strict enforcement~\cite{ablin2022fast}. In its standalone, linear form, Oja’s rule naturally yields solutions constrained to the Stiefel manifold. However, once combined with error-based updates, perfect orthonormality is rarely achieved because the task-specific gradient displaces the solution from the manifold. In this sense, our combined scheme conceptually aligns with “approximate” manifold-based techniques, keeping solutions close to, but not strictly on, the manifold.

From the classical perspective, principal subspace methods like Oja’s rule arise from objectives involving variance maximization or residual minimization. While the hybrid model relaxes strict orthogonality, the residual error remains well-controlled~\cite{karhunen1994representation}. A direct consequence of maintaining a smaller residual error is that layer weights become effectively reversible, allowing upstream signals to be reconstructed simply by transposing the forward weights and reducing the memory cost of storing intermediate activations during training. This property is especially attractive for neuromorphic systems, where local signal reconstruction and limited memory resources are critical. Standard backprop does not inherently guarantee such layer-wise invertibility (see Supplementary note~\ref{sec:reverse}).

Moreover, while integrating Oja’s rule with backprop no longer identifies the exact principal subspace, it still facilitates robust learning. This observed stability sharply contrasts with hierarchical variants of principal subspace analysis, such as Sanger’s rule~\cite{sanger1989optimal}, which tend to falter when coupled with error-based learning (see Supplementary note~\ref{sec:sanger}). Sanger’s rule converges to principal components, whereas Oja’s rule tolerates rotations in the principal subspace, accommodating the dual requirements of error-based learning and residual minimization.

Oja originally proposed his subspace rule as biologically plausible, noting that each synaptic update depends solely on the pre-synaptic input, the post-synaptic response, and an auxiliary feedback variable~\cite{oja1992principal}. Subsequent work, however, argued that the orthonormalization term requires the post-synaptic neuron to account for activity from neighboring neurons~\cite{kuan1991convergence, pehlevan2015hebbian}. Several approaches incorporate additional neural circuitry to plausibly approximate or transmit these lateral signals~\cite{xu1993least}. For instance, Karhunen and Joutsensalo~\cite{karhunen1994representation} introduced a recurrent feedback pathway to unify the required terms. Similarly, F\"{o}ldi\'{a}k~\cite{foldiak1989adaptive} and Pehlevan~\textit{et al.}~\cite{pehlevan2015hebbian} employed Hebbian–anti-Hebbian designs with lateral connections to implement principal subspace updates. Collectively, these models highlight that Oja’s rule can be realized by neural circuitry, provided one accepts the inclusion of auxiliary layers or lateral connections.

We deliberately focused on fully connected architectures, as these models permit a clear interpretation of plasticity rules. While convolutional and attention-based layers have become central to contemporary tasks, their inherent weight-sharing and multi-head structures complicate a straightforward Hebbian interpretation. Extending Oja’s rule and related biologically motivated constraints to complex architectures represents a compelling direction for future work.

In summary, although deceptively simple, Oja’s rule becomes remarkably effective when combined with error-driven learning. Our results highlight a biologically grounded approach that overcomes common training pitfalls without relying on engineered shortcuts. Enhanced stability and improvements over backprop suggest a pivotal shift toward next-generation neural systems. Future work could extend Oja’s stabilizing effect through advanced Hebbian variants or neuromorphic implementations, ultimately advancing the integration of engineered deep learning with biological principles and paving a promising path for neuroscience-inspired artificial intelligence.

\newpage

\section*{Methods}

\subsection*{Models}

We implemented two distinct feedforward neural network architectures to perform the 10-class classification task on the MNIST dataset. Networks were trained online, with one sample presented per iteration (batch size = 1) over a single epoch. Both architectures were initialized via the Xavier method (except as noted in Fig.~\ref{fig:naive}) and employed a softplus activation function with $\beta=10$. Input images with dimensions of $28 \times 28$ pixels (784 input features) were processed by these networks, with the resulting outputs evaluated using cross-entropy loss. The baseline architecture consisted of five layers with dimensions $784$-$170$-$130$-$100$-$70$. A deeper 10-layer architecture was configured with layer sizes $784$-$100$-$90$-$80$-$70$-$60$-$50$-$40$-$30$-$20$. The five-layer architecture was used in experiments presented in Fig.~\ref{fig:depth}a, Fig.~\ref{fig:naive}, and Fig.~\ref{fig:BioOja} (for EMNIST), while the 10-layer architecture was employed for Fig.~\ref{fig:depth}b, Fig.~\ref{fig:residual}, and Fig.~\ref{fig:norm}.

For the memory task (Fig.~\ref{fig:RNN}), we implemented a recurrent neural network (RNN) consisting of a single recurrent layer with 64 hidden units. The RNN received 784-dimensional inputs and produced 10-dimensional outputs, with a hyperbolic tangent activation function and a leak factor ($\alpha=0.8$) during simulations. Each trial spanned 50 timesteps, with the initial 5 timesteps for sample presentation, followed by a delay period. The network's prediction was evaluated at the final timestep using cross-entropy loss. This RNN was trained online on the MNIST dataset for 18,000 iterations.

For meta-learning experiments (Fig.~\ref{fig:BioOja}), the five-layer feedforward architecture described above was adapted for 5-class classification tasks using the EMNIST dataset, with models trained on 100 samples per task (20 samples per class).

All experiments were conducted over 20 independent runs, with figures displaying the mean values and error bars representing 98\% confidence intervals computed via bootstrapping over 500 samples.

\subsection*{Optimal linear mapping}
We derived optimal linear mappings between adjacent layers to assess the extent to which layer-wise activations preserve upstream information. Specifically, we identified a linear transformation for each layer $\ell$ that minimizes the reconstruction error between the pre-synaptic activations $\mathbf{y}_{\ell-1}$ and the transformed post-synaptic activations $\hat{\mathbf{y}}_{\ell-1}$,
\begin{equation}
    \min_{\boldsymbol{R}_{\ell,\ell-1}} \; \|\mathbf{y}_{\ell-1} - \hat{\mathbf{y}}_{\ell-1}\|^2, \quad \text{where} \quad \hat{\mathbf{y}}_{\ell-1}=\boldsymbol{R}_{\ell,\ell-1}\,\mathbf{y}_{\ell}
\end{equation}
Once the training converged, we computed the optimal mappings using a closed-form solution based on the test dataset,
\begin{equation}
    \boldsymbol{R}_{\ell,\ell-1} = \mathbf{y}^\top_{\ell-1}\,\mathbf{y}_{\ell}\,\left(\mathbf{y}^\top_{\ell}\,\mathbf{y}_{\ell}\right)^{-1}.
\end{equation}
After obtaining $\boldsymbol{R}_{\ell,\ell-1}$ for each pair of consecutive layers, we recursively reconstructed the upstream activations. Beginning from the final hidden layer ($\ell = L-1$), we propagated the reconstruction backward through the network using \begin{equation}
    \hat{\mathbf{y}}_{\ell-1} = \boldsymbol{R}_{\ell,\ell-1}\,\hat{\mathbf{y}}_{\ell}, 
    \quad
    \text{where}
    \quad
    \hat{\mathbf{y}}_{L-1} = \mathbf{y}_{L-1}
    \label{eq:fwd_pass}
\end{equation}
This recursive inversion produces a reconstruction of the input $\hat{\mathbf{y}}_0$.

\subsection*{Meta-learning plasticity}

Meta-learning consists of two nested phases: adaptation and meta-optimization. First, the adaptation phase learns the network's forward and backward pathways, $\boldsymbol{W}$ and $\boldsymbol{B}$, via the plasticity rule $\mathcal{F}(\boldsymbol{\Theta})$. Next, the meta-optimization phase adjusts the meta-parameters $\boldsymbol{\Theta}$.

The meta-training dataset consists of $\mathcal{E}$ tasks $\mathcal{T}^{(\varepsilon)}$, where each task trains one meta-iteration (episode). Each task contains $K$ training examples $\mathcal{T}^{(\varepsilon)}_{\text{train}}$ and $Q$ query examples $\mathcal{T}^{(\varepsilon)}_{\text{query}}$ per class. The training examples update $\boldsymbol{W}$ and $\boldsymbol{B}$, whereas the query examples guide the optimization of $\boldsymbol{\Theta}$. For our experiments (Fig.~\ref{fig:BioOja}), we set $K=20$ and $Q=10$ per class and performed 5-class classification on tasks drawn from the EMNIST dataset.

At each meta-iteration (episode), $\boldsymbol{W}$ and $\boldsymbol{B}$ are reinitialized separately via the Xavier scheme. Then, adaptation is performed online, one data point at a time from $\mathcal{T}^{(\varepsilon)}_{\text{train}}$, using a plasticity rule defined as a linear combination of $R$ candidate terms $\{\mathcal{F}^r\}_{0\leq r\leq R-1}$:
\begin{equation}
    \mathcal{F} (\boldsymbol{\Theta}) = \sum_{r=0}^{R-1} \theta_r \mathcal{F}^{r}.
    \label{eq:linear}
\end{equation}
where $\boldsymbol{\Theta}=\{\theta_r \mid 0\leq r\leq R-1\}$ are shared across layers. At every adaptation step, the loss is computed using the cross-entropy function.

Subsequently, at the end of each episode, $\boldsymbol{\Theta}$ is updated using cross-entropy loss for meta-optimization. Although alternative methods (e.g., evolutionary algorithms) exist, we chose the gradient-based ADAM optimizer~\cite{kingma2014adam}. Consequently, the network had to be twice differentiable to accommodate gradient computation through the unrolled adaptation loops. Therefore, we selected the softplus activation function, a continuously differentiable approximation of ReLU, for all models. In our experiments (Fig.~\ref{fig:BioOja}b-d), the coefficient $\theta_0$ was initialized to $10^{-3}$ and $\theta_{1}$, $\theta_{2}$, $\theta_{3}$, $\theta_{4}$, $\theta_{5}$, and $\theta_{6}$ were set to zero. We used a meta-learning rate of $2\times10^{-4}$.

\subsection*{Backprop through time}

In the present framework, the recurrent dynamics update the hidden state $\mathbf{h}_t$ via a convex combination of its previous state and a newly computed pre-activation term:
\begin{equation}
    \mathbf{h}_{t+1} \;=\; \alpha\,\mathbf{h}_{t} \;+\; (1-\alpha) \,\Bigl(\boldsymbol{W}_h\,\sigma(\mathbf{h}_{t}) \;+\;\boldsymbol{W}_x\,\mathbf{x}_{t+1}\Bigr),
\end{equation}
Here, $\sigma$ denotes the nonlinear activation function, while $\boldsymbol{W}_h$ and $\boldsymbol{W}_x$ denote the recurrent and input weight matrices, respectively. A readout is performed at the final time step $T$.
The parameter $\alpha\in (0,1]$ is a leak coefficient. 
Taking $\alpha=0$ produces vanilla RNN models commonly used in machine learning while taking $\alpha$ close to $1$ produces a forward Euler approximation to continuous-time RNN models widely used in computational neuroscience~\cite{rosenbaum2024modeling}. In principal, non-zero values of $\alpha$ introduce memory into the state, $\mathbf{h}_{t}$. However, the memory trace decays like $\alpha^t$ where $t$ is the number of time steps. Hence, this form of memory alone is not sufficient for our task in which we use $\alpha=0.8$ with a delay period of $t=45$ time steps.

To train our baseline RNN, we employ Backprop Through Time (BPTT), which unfolds the network over time to compute gradients for updating its weights. The network output at the final time step is computed as
\begin{equation}
    \tilde{\mathbf{y}}_T \;=\; \boldsymbol{W}_o \,\sigma\bigl(\mathbf{h}_T\bigr),
\end{equation}
with $\boldsymbol{W}_o$ representing the output weight matrix. Conventional gradient-based training via BPTT computes the gradient of the loss $\mathcal{L}$ with respect to $\boldsymbol{W}_h$. The update of $\boldsymbol{W}_h$ under standard BPTT takes the form
\begin{equation}
\Delta \boldsymbol{W}_h \;=\; \theta \;\sum_{t=1}^{T-1}\mathbf{e}_{t+1}\,\sigma(\mathbf{h}_t)^\top,
\end{equation}
where $\theta$ is the learning rate and $\mathbf{e}_t$ denotes the backpropagated error, defined recursively as
\begin{equation}
\mathbf{e}_t \;=\; \boldsymbol{W}_h^\top \,\mathbf{e}_{t+1} \,\odot\, \sigma'\bigl(\mathbf{h}_t\bigr), \quad
\mathbf{e}_T \;=\;\frac{\partial \mathcal{L}}{\partial \mathbf{h}_T},
\end{equation}
with $\odot$ indicating elementwise multiplication.

\subsection*{Quantifying stimulus-driven neural variance}

To assess the impact of different plasticity rules on the information encoded by individual neurons about the presented digit in a recurrent network model, we performed an analysis of variance (ANOVA) on hidden states across trials under sample-dependent conditions. ANOVA partitions the total variance into a portion explained by the factor (between-group) and a residual portion (within-group/error). This procedure enabled us to compute the proportion of explained variance (PEV), quantifying how strongly each neuron's activity depends on the stimulus condition.

In this analysis, the categorical variable \(\text{C(sample)}\) represents the digit identity in each image, serving as the experimental condition. The bias-corrected effect size \(\omega^2\)~\cite{olejnik2003generalized} is then defined as
\begin{equation}
    \omega^2 = \frac{\text{SS}_{\text{factor}} - \bigl(\text{df}_{\text{factor}} \times \text{MS}_{\text{error}}\bigr)}{\text{SS}_{\text{total}} + \text{MS}_{\text{error}}},
\end{equation}
where \(\text{SS}_{\text{factor}}\) is the sum of squared deviations of each class mean from the overall mean, weighted by the number of observations in that class:
\begin{equation}
    \text{SS}_{\text{factor}} = \sum_{g=1}^{G} n_g \bigl(\bar{x}_g - \bar{x}\bigr)^2.
\end{equation}
Here, \(\bar{x}\) is the overall mean activation across all observations; \(\bar{x}_g\) denotes the mean activation of observations in digit class \(g\); and \(n_g\) is the number of observations belonging to class \(g\).

The total sum of squares is given by
\begin{equation}
    SS_{\text{total}} = \sum_{g=1}^{G} \sum_{i \in g} \bigl(x_i - \bar{x}\bigr)^2,
\end{equation}
where \(x_i\) denotes the neural activation of the \(i\)-th observation. Meanwhile, the error sum of squares,
\begin{equation}
    SS_{\text{error}} = \sum_{g=1}^{G} \sum_{i \in g} \bigl(x_i - \bar{x}_g\bigr)^2,
\end{equation}
captures the variability not explained by the factor. The mean square error is
\begin{equation}
    \text{MS}_{\text{error}} = \frac{\text{SS}_{\text{error}}}{\text{df}_{\text{error}}},
\end{equation}
and \(\text{df}_{\text{factor}}\) is the degrees of freedom for the digit-identity factor, typically one less than the number of classes.

\section*{Acknowledgments}
This work was supported by Air Force Office of Scientific Research (AFOSR) grant FA9550-21-1-0223 (N.S.T., M.A.M., R.R.), National Science Foundation (NSF) Neuronex award DBI-1707400 (R.R.), and by the Swartz Foundation Fellowship for Theory in Neuroscience (N.S.T.).

\bibliographystyle{unsrtFirstInit}
\bibliography{references}

\beginsupplement

\setcounter{subsection}{0}
\renewcommand{\thesubsection}{S\arabic{subsection}}

\section*{Supplementary Material}

\subsection{Mechanistic interplay of Oja's rule and backprop \label{sec:supp_oja_vs_bp}}

To demonstrate how Oja's rule improves learning in a network initialized with unscaled weights, we examined the information flow of the network presented in Fig.~\ref{fig:naive}. Specifically, we tracked (i) the step sizes for backprop and Oja's term in $\mathcal{F}^{\text{Oja}}$ (equation~(\ref{eq:BPOja})), (ii) the residual error $\mathcal{J}$ (equation~(\ref{eq:res_err})), and (iii) norm of the backprop error $\mathbf{e}$ (equation~(\ref{eq:e_l_backprop})) across layers. 

\begin{figure}[H]
    \setlength{\unitlength}{0.01\textwidth}
    \begin{picture}(98,59.5)
        \put(1,58){\footnotesize a}
        \put(2,29){\includegraphics[width=0.46\textwidth]{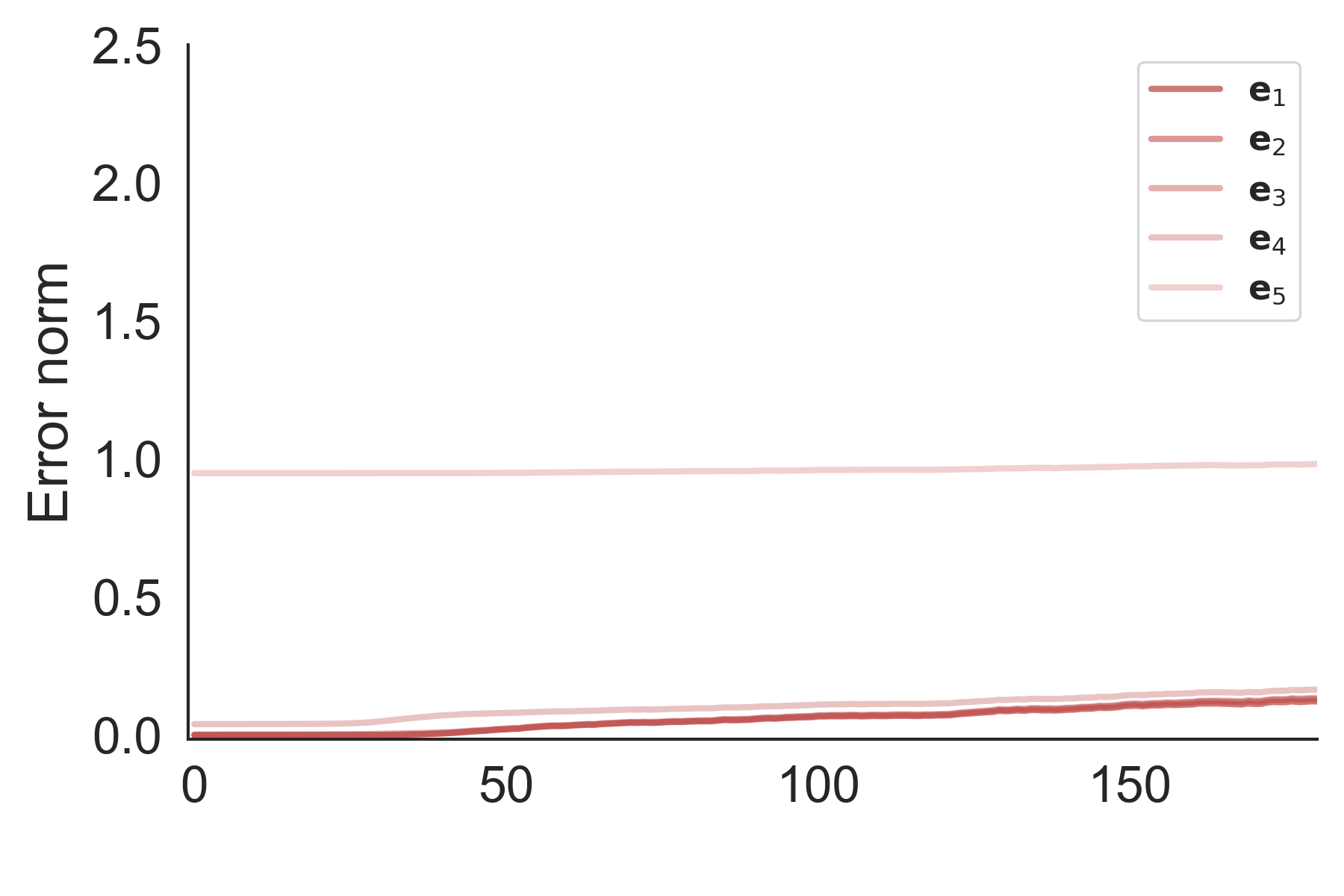}}
        \put(51,58){\footnotesize b} 
        \put(52,29){\includegraphics[width=0.46\textwidth]{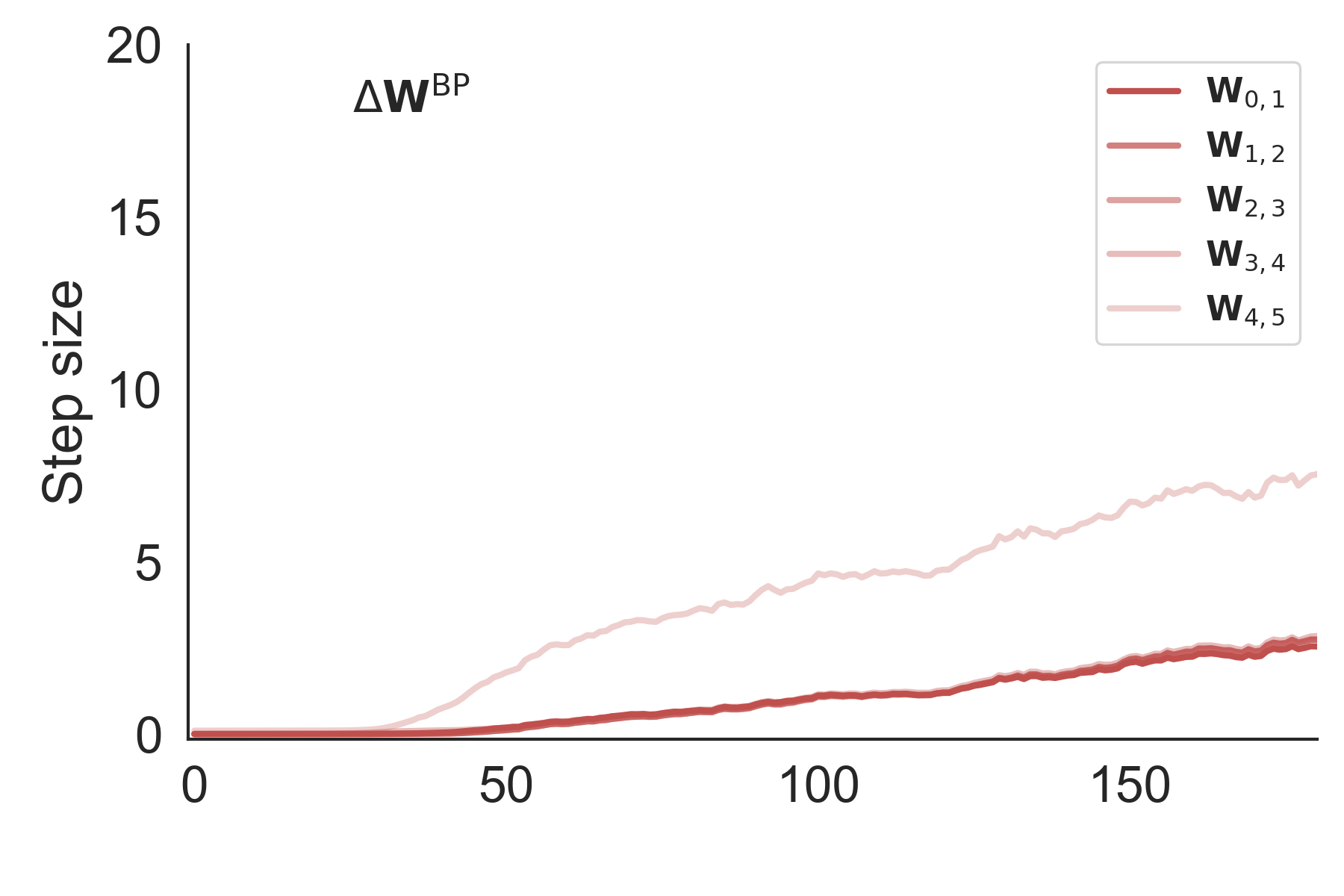}}
        
        \put(1,29){\footnotesize c} 
        \put(2,0){\includegraphics[width=0.46\textwidth]{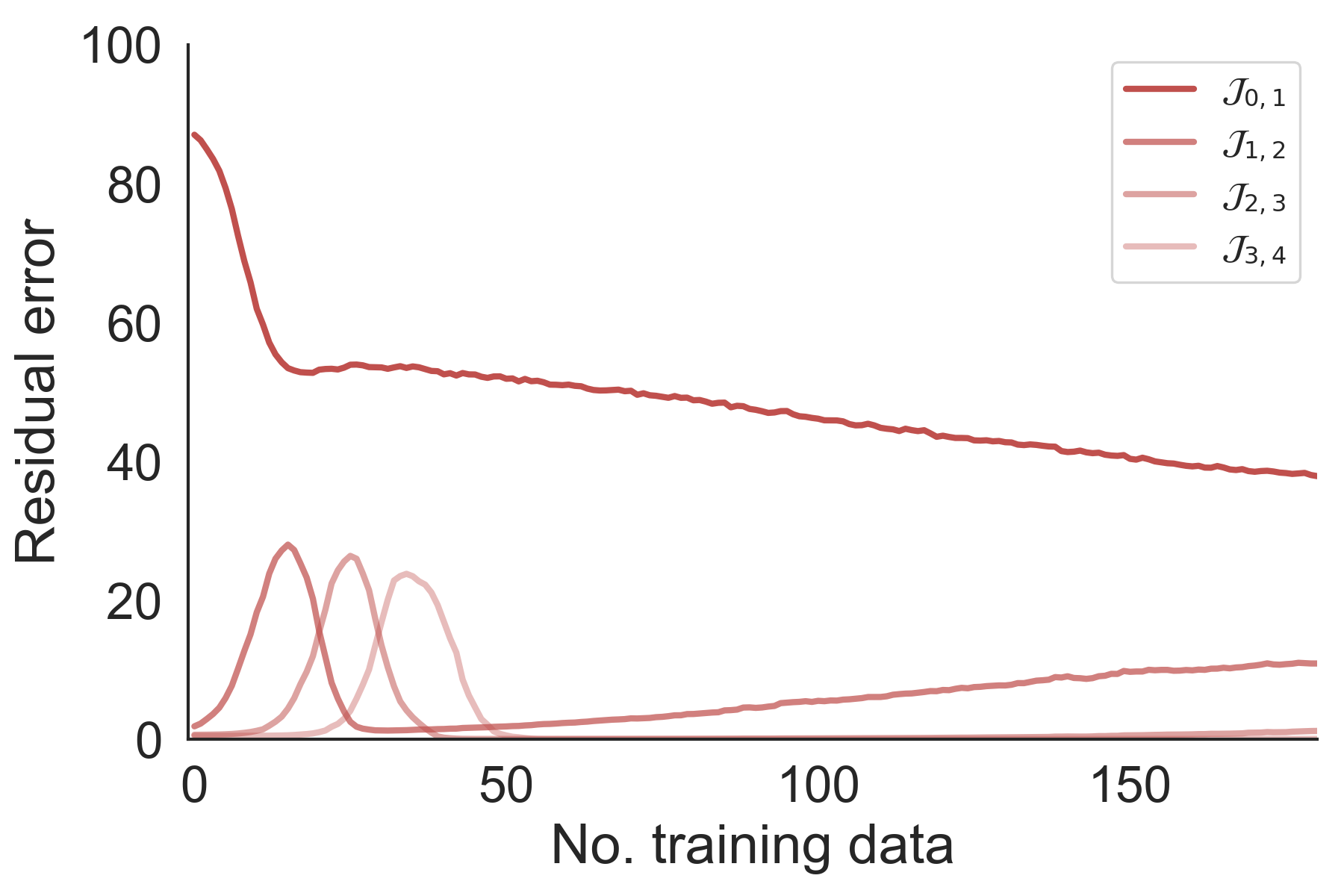}} 
        \put(51,29){\footnotesize d}
        \put(52,0){\includegraphics[width=0.46\textwidth]{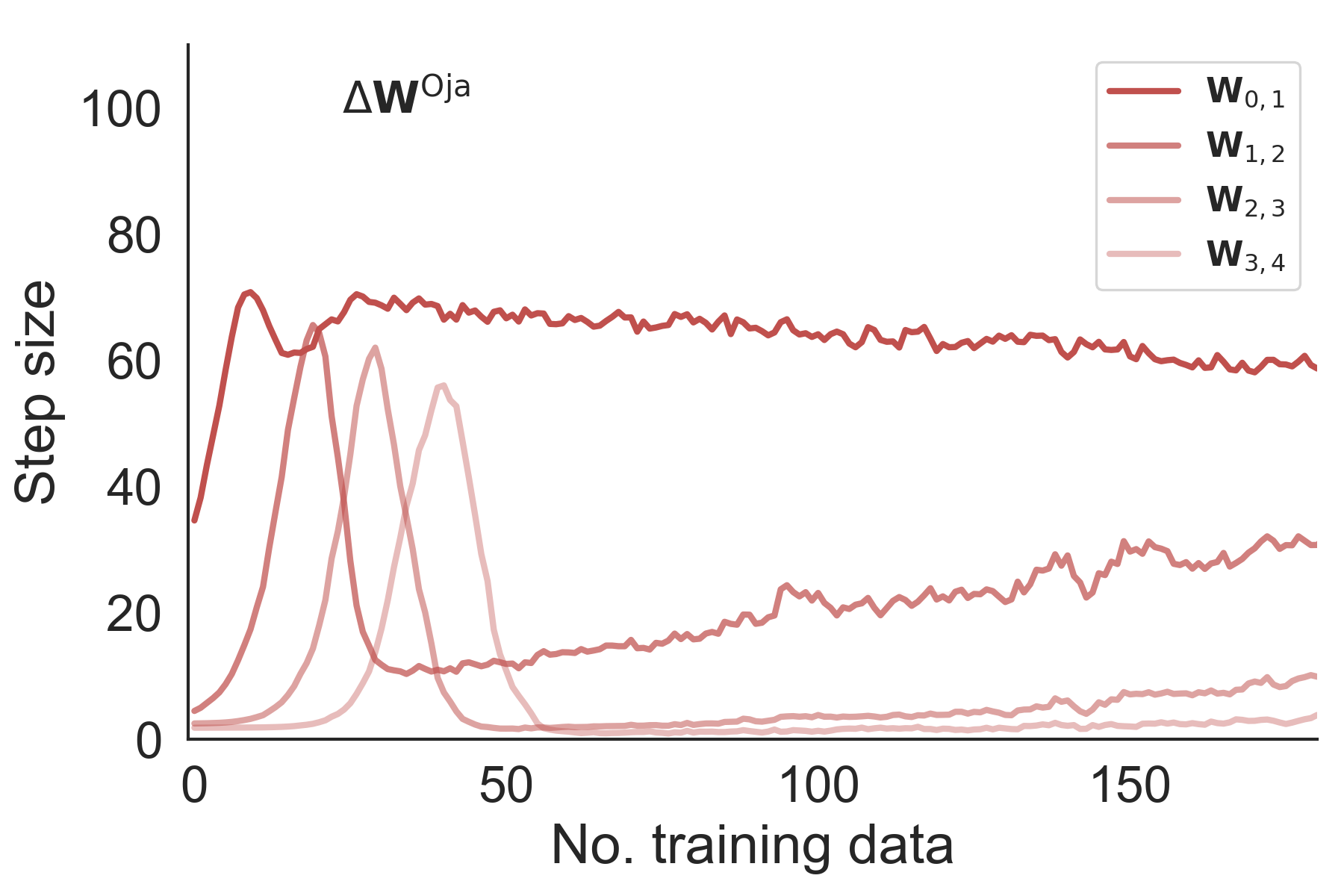}} 
    \end{picture}
    \caption{\textbf{Oja's rule mitigates vanishing signals.} Optimization elements in the early training stage (180 iterations) under naive initialization using $\mathcal{F}^{\text{Oja}}$ (equation~(\ref{eq:BPOja})) for a 5-layer network.
    \textbf{a}, Norm of backprop modulatory errors $\|\mathbf{e}_{\ell}\|$ (equation~(\ref{eq:e_l_backprop})) and
    \textbf{b}, magnitude of the backprop component in $\mathcal{F}^{\text{Oja}}$ are computed across network layers $\ell=0,\dots, 5$.
    \textbf{c}, Residual error $\mathcal{J}_{\ell-1, \ell}$ (equation~(\ref{eq:res_err})) and
    \textbf{d}, Oja’s term step size are computed within feature-extraction layers $\ell=0,\dots, 4$. The last layer is omitted from Oja's term plots in panels \textbf{c} and \textbf{d} as it does not affect learning in the prediction layer by design. Darker to lighter hues indicate shallower to deeper layers. Each curve shows the mean of 20 test runs.}
    \label{fig:step}
\end{figure}

When the network is initialized with unscaled weights, forward activations and backward modulatory signals are severely diminished (Fig.~\ref{fig:step}a). Consequently, the backprop step size, which relies on pre-synaptic activation and post-synaptic error, remains minuscule throughout the network (Fig.~\ref{fig:step}b). In contrast, the nontrivial pre-synaptic activations in the first layer yield a sizeable residual error $\mathcal{J}_{0, 1}$ (Fig.~\ref{fig:step}c). In response, Oja's term increases, regulating post-synaptic activation norm and reducing the discrepancy (Fig.~\ref{fig:step}d). As the first layer's residual subsides, the second layer experiences an increase. Though seemingly paradoxical, the initially small error in deeper layers merely reflects vanishing signals (i.e., ``garbage in, garbage out''). Once meaningful activations reach this layer, Oja's step size rises in response to the elevated residual. This pattern recurs layer by layer, establishing a ``chain reaction'' wherein elevated residual triggers Oja's rule to stabilize layers and permit informative signals to propagate to deeper ones.

By contrast, backprop step sizes still remain effectively zero throughout the layers. While the last layer always has a non-zero error $\mathbf{e}_5$ (Fig.~\ref{fig:step}a), its ability to produce a substantial update step is hampered until the preceding activation $\mathbf{y}_4$ moves away from near-zero values. Once Oja's step size escalates in the layer before output, its post-synaptic activations amplify, and the final layer's backprop updates gain traction. By the time backprop becomes effective in the last layer, the upstream layers have shifted away from their initial states, enabling gradient-based learning across the network.

Collectively, these observations underscore how Oja's rule offsets the pitfalls of naive weight scaling by progressively rescaling and stabilizing layer activations. This effect obviates the necessity of fine-tuned initial weight distributions or protracted pre-training routines.

\subsection{Activation norm preservation in linear layers under Oja's rule \label{sec:norm_proof}}
Consider the linear forward equation:
\begin{equation}
    \mathbf{y}_{\ell} = \boldsymbol{W}_{\ell-1, \ell} \mathbf{y}_{\ell-1}.
\end{equation}
In the linear case, Oja's rule minimizes the objective:
\begin{equation}
    \mathcal{J} = \frac{1}{T} \sum_{i=1}^{T} \left\| \mathbf{y}_{\ell-1}^{(i)} - \boldsymbol{W}_{\ell-1, \ell}^\top \left( \boldsymbol{W}_{\ell-1, \ell} \mathbf{y}_{\ell-1}^{(i)} \right) \right\|^2.
    \label{eq:res_err_linear}
\end{equation}
Thus if the minimum at $\mathcal{J}=0$ is realized then:
\begin{equation}
\begin{aligned}
    \|\mathbf{y}_{\ell-1} \|^2 &= \mathbf{y}_{\ell-1}^\top \mathbf{y}_{\ell-1} \\
    &= \mathbf{y}_{\ell-1}^\top \left[ \boldsymbol{W}_{\ell-1, \ell}^\top \left( \boldsymbol{W}_{\ell-1, \ell} \mathbf{y}_{\ell-1} \right) \right] \\
    &= \left( \boldsymbol{W}_{\ell-1, \ell} \mathbf{y}_{\ell-1} \right)^\top \left( \boldsymbol{W}_{\ell-1, \ell} \mathbf{y}_{\ell-1} \right) \\
    &= \mathbf{y}_{\ell}^\top \mathbf{y}_{\ell} \\
    &= \| \mathbf{y}_{\ell} \|^2.
\end{aligned}
\label{eq:norm_proof}
\end{equation}
The derivation equation~(\ref{eq:norm_proof}) shows that once the objective $\mathcal{J}$ (equation~(\ref{eq:res_err_linear})) is driven to zero, the Euclidean norm of the activations remains constant across layers. 

\subsection{Layer-wise Preservation of Variance \label{sec:ve}}

In the main text, we focus on how well the first hidden layer, under backprop (equation~(\ref{eq:BP})) versus hybrid backprop plus Oja's update (equation~(\ref{eq:BPOja})), preserves variance in both the activations themselves and their linear reconstructions (Fig.~\ref{fig:residual}d). Here, we extend that analysis to deeper layers to demonstrate that Oja's rule consistently promotes broader variance retention across the network depth.

\begin{figure}[H]
    \setlength{\unitlength}{0.01\textwidth}
    \begin{picture}(98,31)
        \put(1,29.5){\footnotesize a}
        \put(2,0){\includegraphics[width=0.46\textwidth]{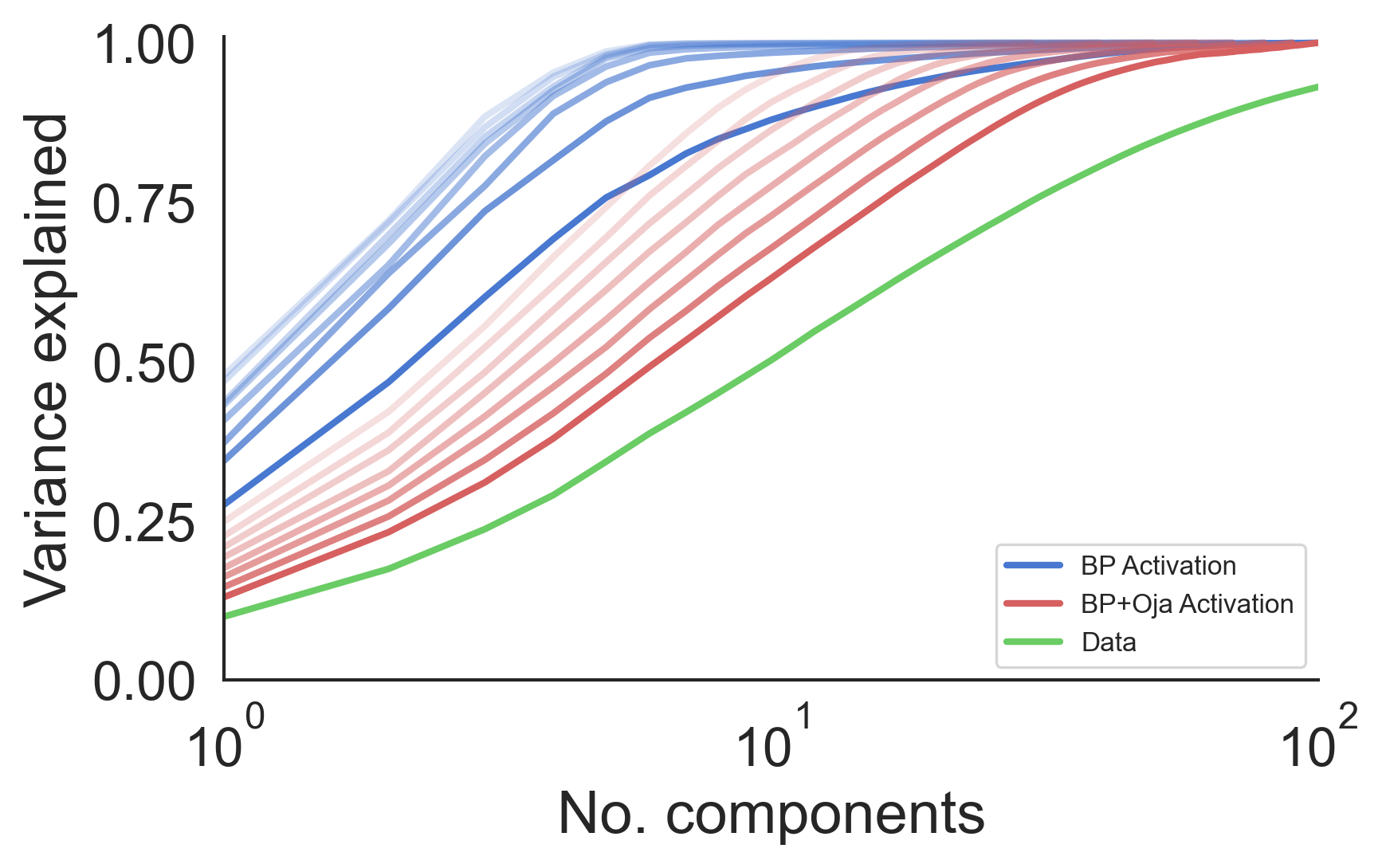}}
        \put(51,29.5){\footnotesize b}
        \put(52,0){\includegraphics[width=0.46\textwidth]{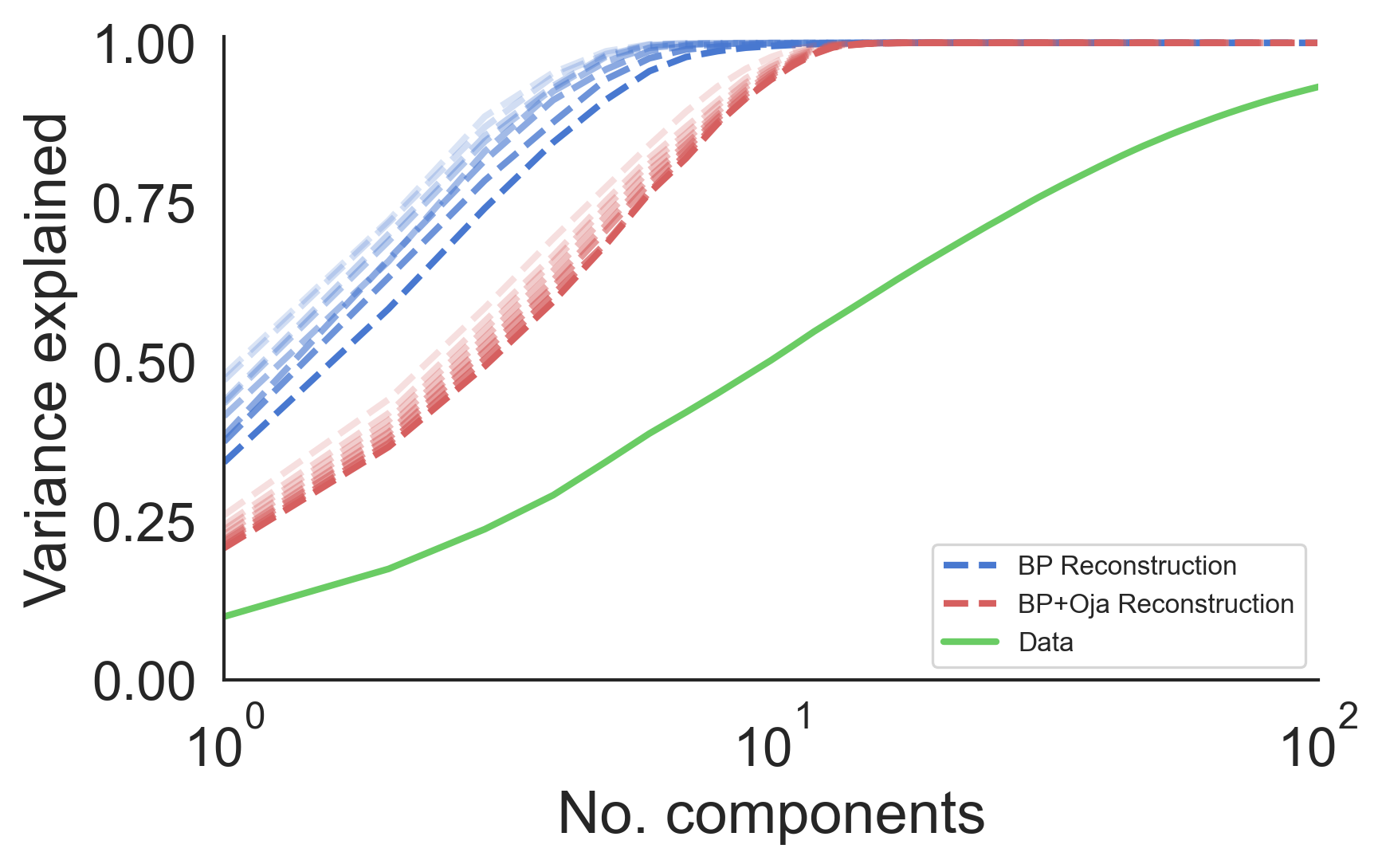}}
    \end{picture}
    \caption{\textbf{Oja's rule retains broader variance across layers.}
    \textbf{a}, For each layer ($\ell=1$ through 8), we computed the cumulative variance explained of activations $\mathbf{y}_{\ell}$ by principal component analysis (PCA) for backprop (BP; blue curves) and backprop with Oja's rule (BP+Oja; red curves). The green curve shows the corresponding PCA on the input data (for reference). In each deeper layer, the BP+Oja model exhibits a more uniform distribution of variance across principal components, remaining closer to the input data's green curve. 
    \textbf{b}, Each layer $\hat{\mathbf{y}}_{\ell}$ is reconstructed by an optimal linear map from the final hidden layer, and then PCA is performed (BP: dashed blue; BP+Oja: dashed red). Consistent with panel a, the BP+Oja reconstructions remain closer to the input reference, suggesting that Oja's term promotes more balanced variance retention through successive layers.}
    \label{fig:EV_all}
\end{figure}

Figure~\ref{fig:EV_all}a shows the fraction of variance (y-axis) explained as a function of the number of principal components (x-axis) for each hidden layer. Across increasing layer depth (darker to lighter hues of each color set), the hybrid rule's curves (red) retain variance in a more balanced manner than standard backprop (blue). Notably, the BP+Oja curves lie closer to the input data's reference curve (green) in every layer.

To quantify how faithfully each layer's representation could be reconstructed from the final hidden layer, we applied an optimal linear mapping (as detailed in the Methods of the main text). As shown in Fig.~\ref{fig:EV_all}b, the BP+Oja reconstructions (dashed red lines) exhibit explained-variance curves that remain closer to the input reference compared to standard BP (dashed blue lines). This pattern is observed consistently through deeper layers, aligning with the broader variance distributions in the activations themselves.

In summary, these supplementary results reinforce our primary findings that combining error-based updates with Oja's rule helps maintain a broader variance profile across hidden layers. The greater alignment with the input PCA structure (green curves) indicates that the variance distribution is less compressed while retaining the same dimensionality.

\subsection{Rapid norm re-scaling and maintenance with Oja's rule \label{sec:norm_plot}}

\begin{figure}[H]
    \setlength{\unitlength}{0.01\textwidth}
    \begin{picture}(98,111)
        \put(1,110){\footnotesize a}
        \put(2,81){\includegraphics[width=0.46\textwidth]{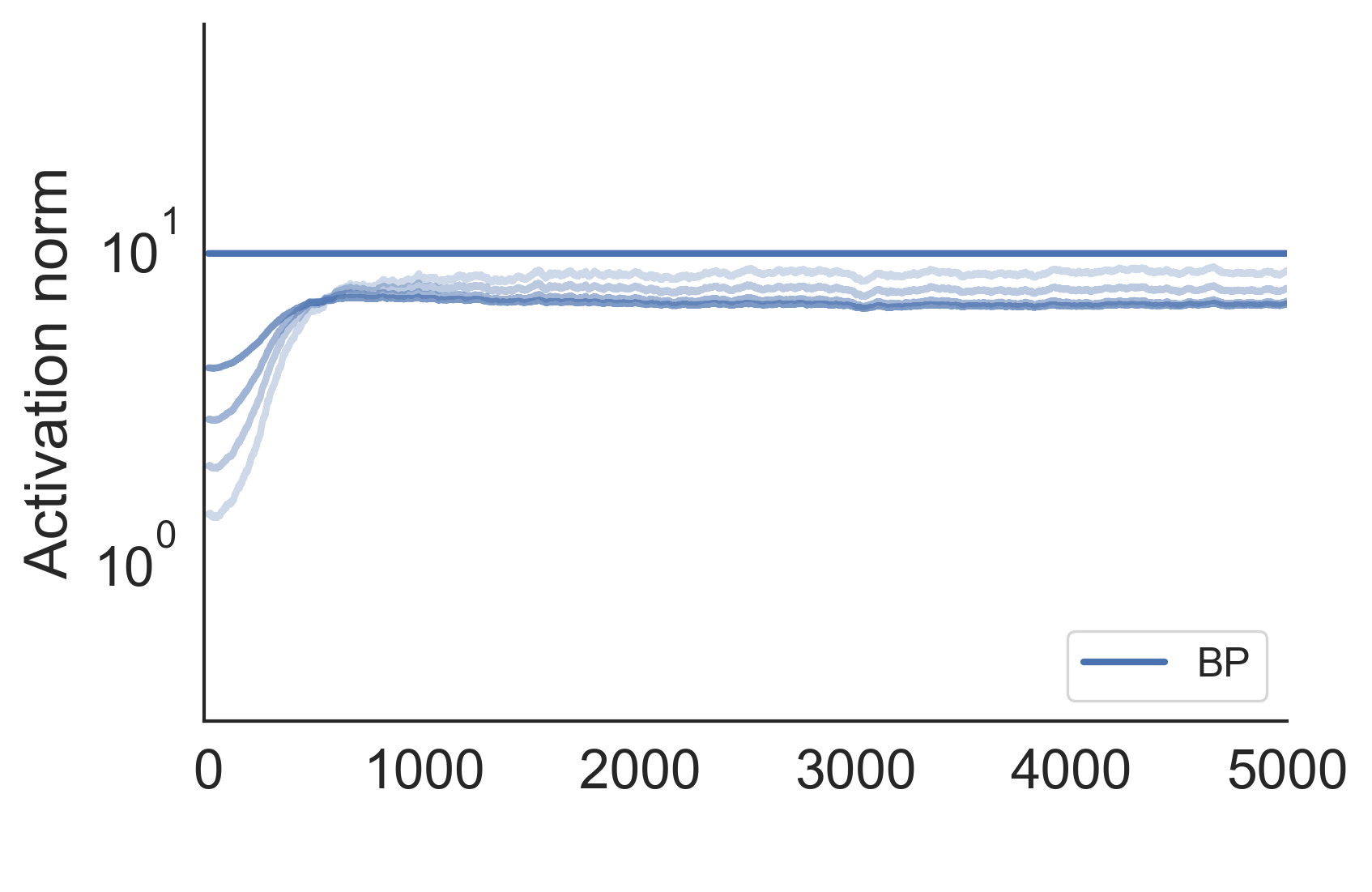}}
        \put(51,110){\footnotesize b}
        \put(52,81){\includegraphics[width=0.46\textwidth]{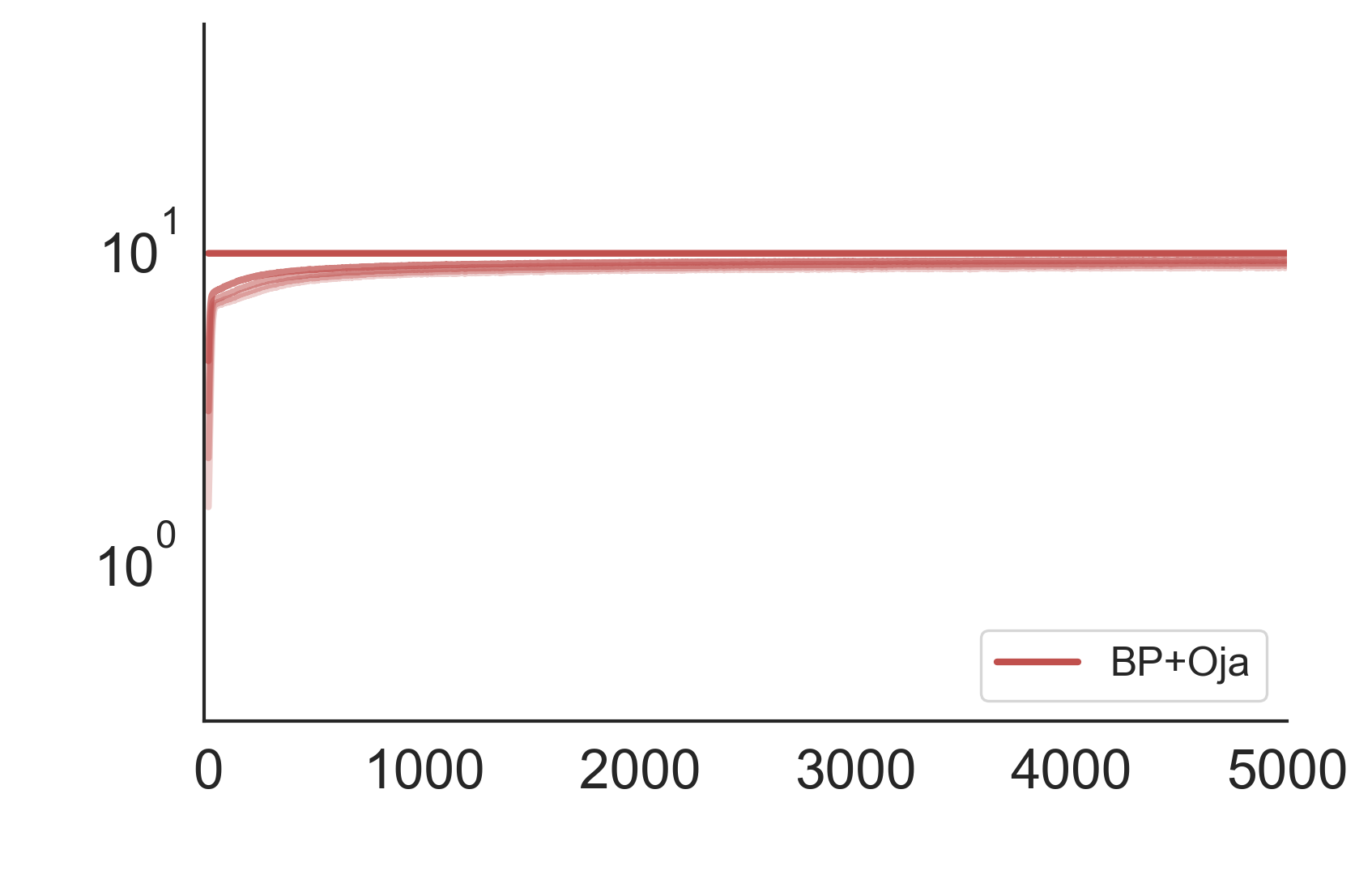}}
        
        \put(1,83){\footnotesize c}
        \put(2,54){\includegraphics[width=0.46\textwidth]{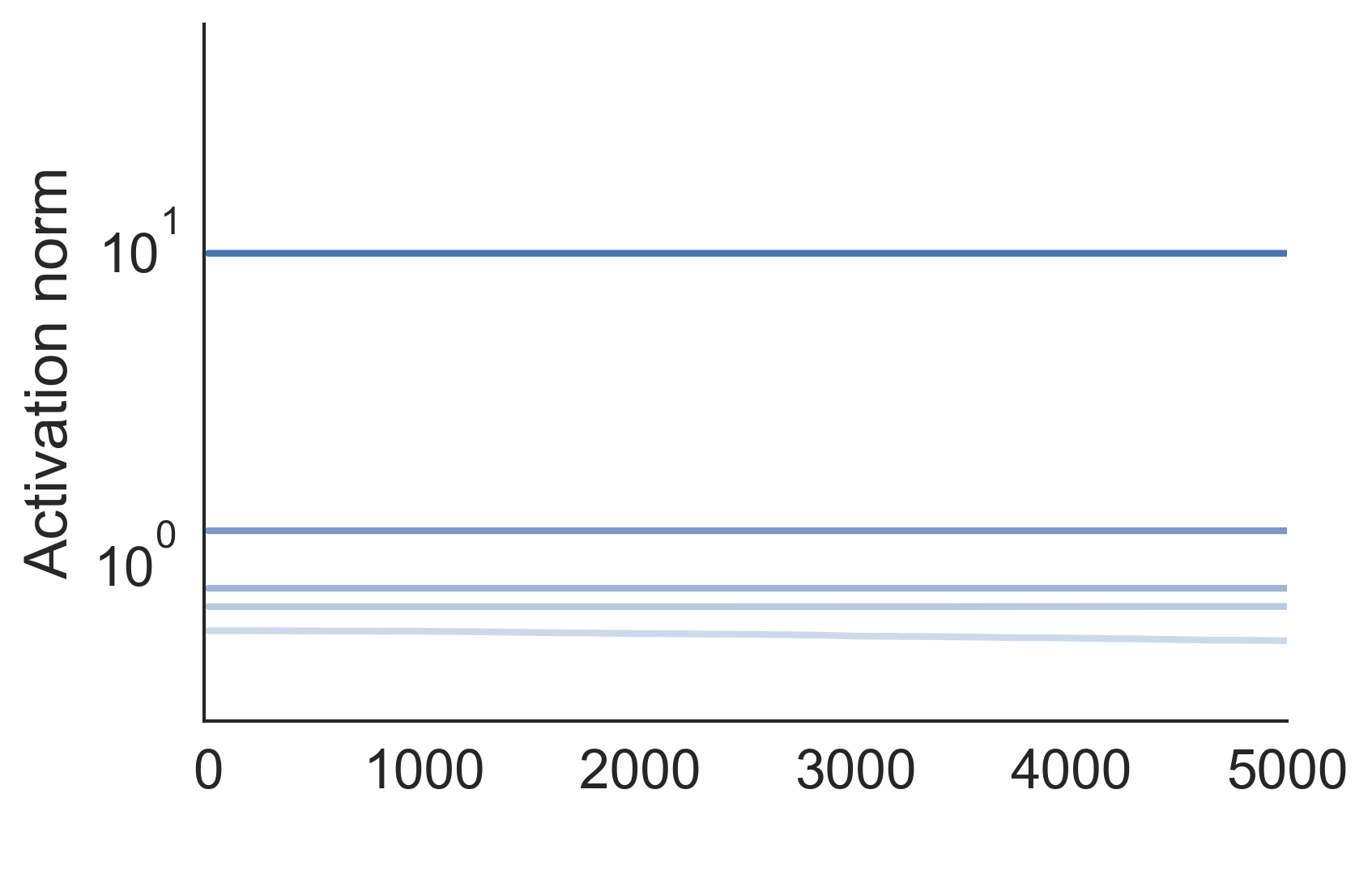}}
        \put(51,83){\footnotesize d}
        \put(52,54){\includegraphics[width=0.46\textwidth]{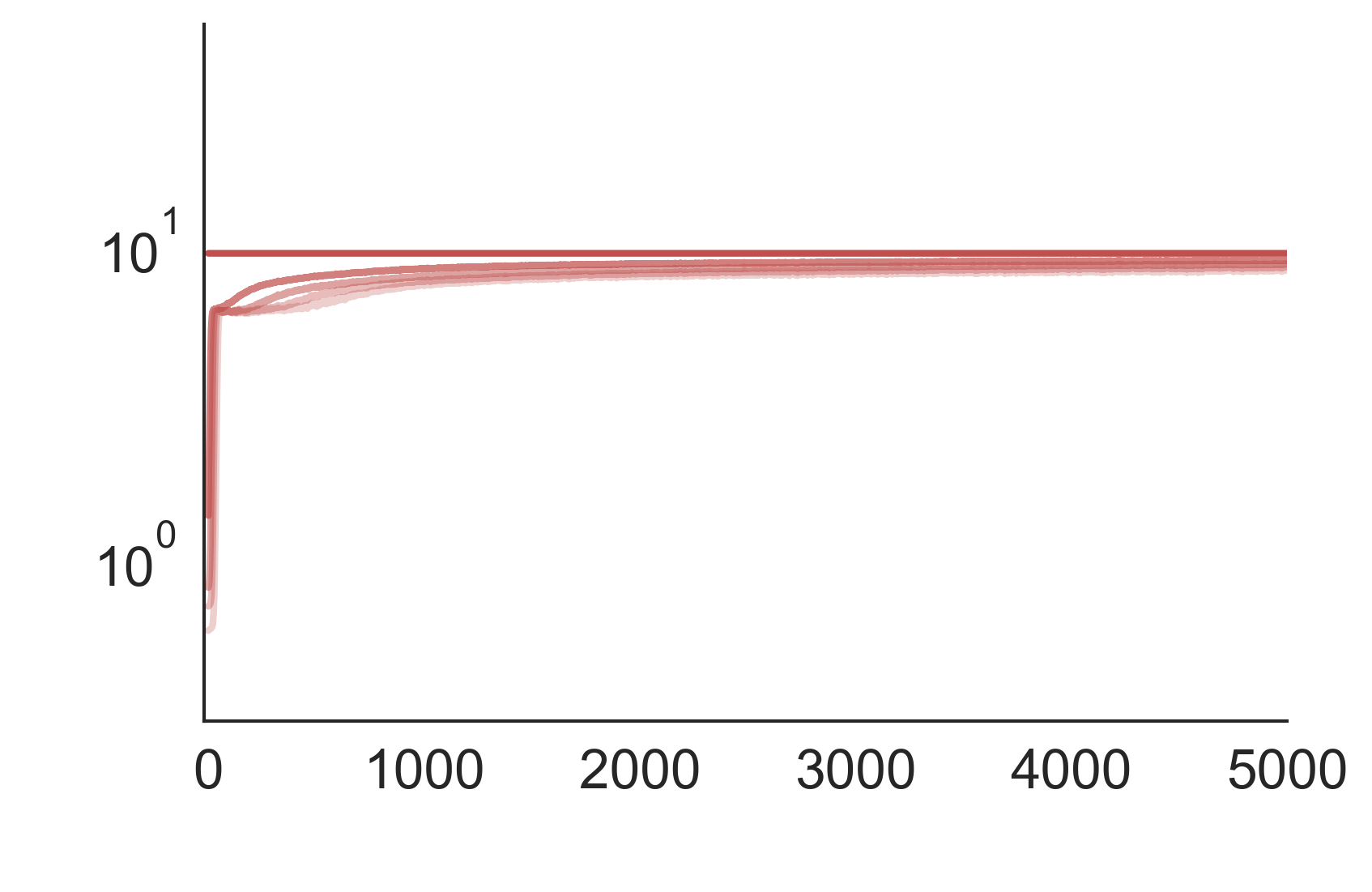}}
        
        \put(1,56){\footnotesize e}
        \put(2,27){\includegraphics[width=0.46\textwidth]{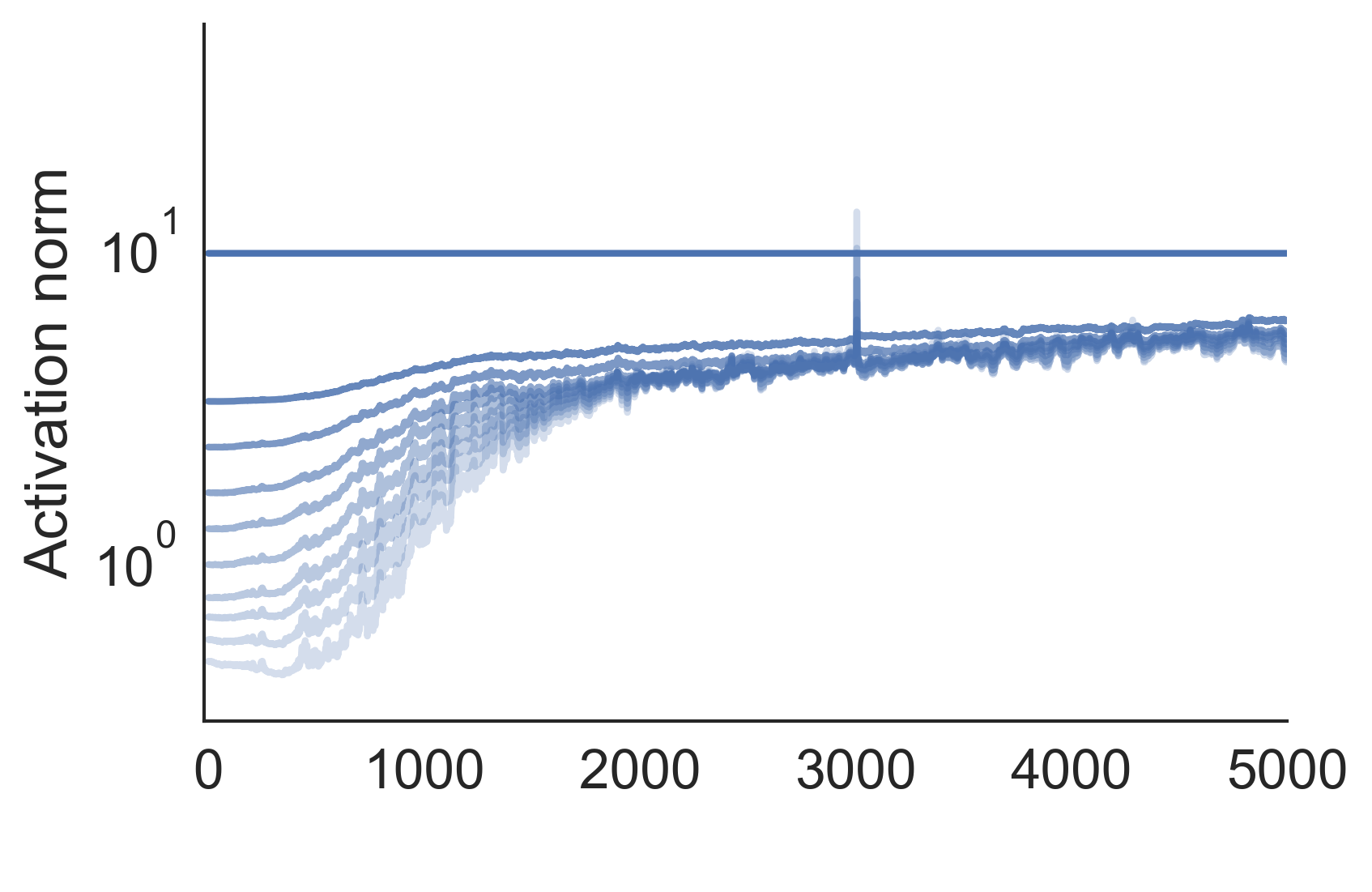}}
        \put(51,56){\footnotesize f} 
        \put(52,27){\includegraphics[width=0.46\textwidth]{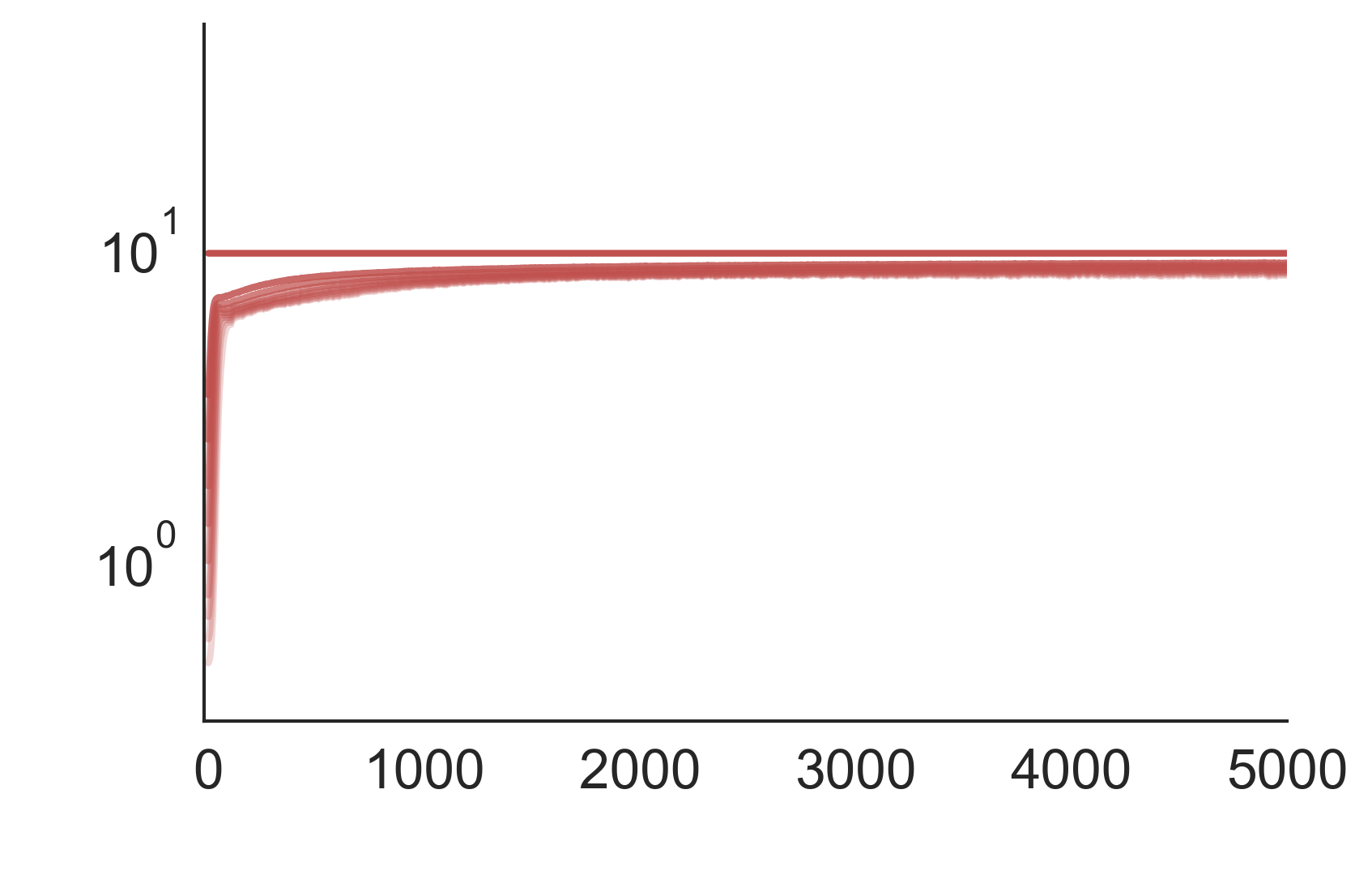}}
        
        \put(1,29){\footnotesize g} 
        \put(2,0){\includegraphics[width=0.46\textwidth]{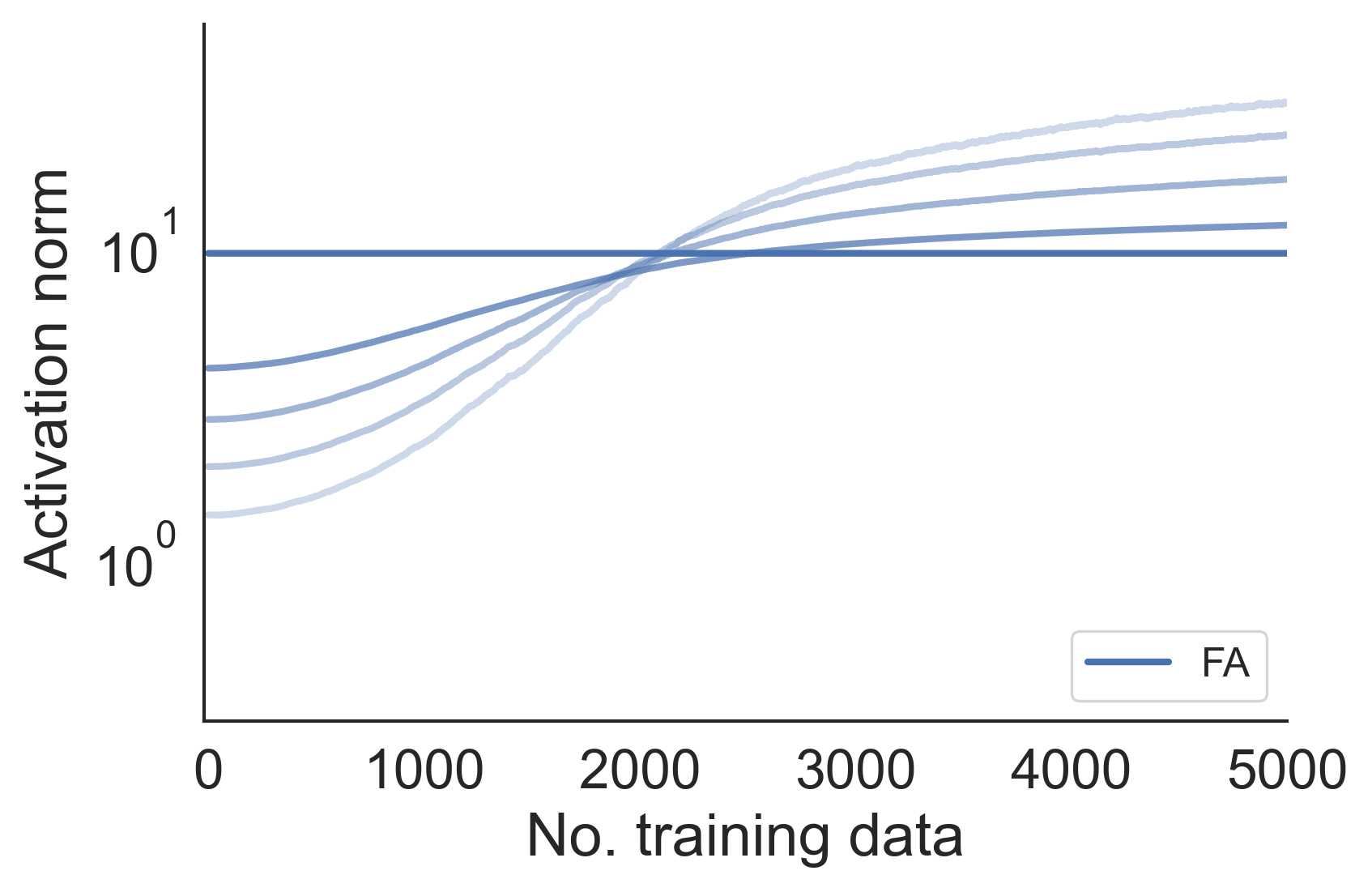}} 
        \put(51,29){\footnotesize h}
        \put(52,0){\includegraphics[width=0.46\textwidth]{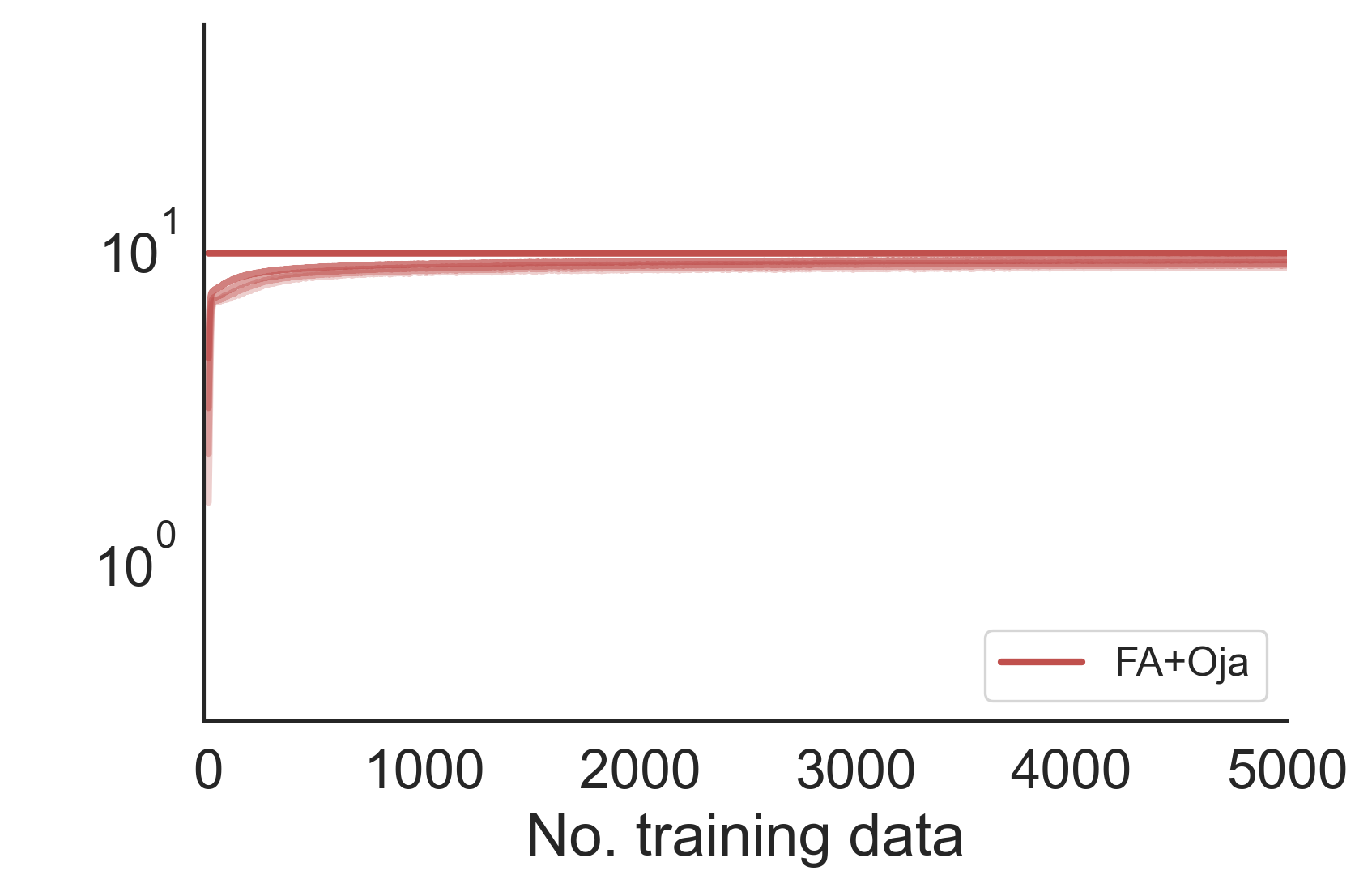}} 
    \end{picture}
    \caption{\textbf{Activation norms remain stable throughout training with Oja's rule.} Activation norms are compared throughout online training in \textbf{a}, \textbf{c}, \textbf{e}, neural networks trained by standard backprop (BP) versus \textbf{b}, \textbf{d}, \textbf{f}, backprop with Oja's rule (BP+Oja), and in \textbf{g}, network trained using feedback alignment (FA) versus \textbf{h}, FA with Oja's rule (FA+Oja). Network variants include \textbf{a}, \textbf{b}, a standard 5-layer architecture, \textbf{c}, \textbf{d}, a 5-layer architecture with naive initialization, \textbf{e}, \textbf{f}, a deeper 10-layer architecture, and \textbf{g}, \textbf{h}, a feedback alignment-based model. The color gradient within each set indicates increasing layer depth (darker to lighter hues).}
    \label{fig:sup_y_norms}
\end{figure}

In Figs.~4 and~6a, we presented activation norms after model convergence. Here, we expand on those findings by illustrating the evolution of activation norms over the course of online training. Figures~\ref{fig:sup_y_norms}a–f depict comparisons between standard backprop (BP) and backprop augmented with Oja's rule (BP+Oja). Figures~\ref{fig:sup_y_norms}g–h contrast random feedback alignment (FA) against FA with Oja's rule (FA+Oja). The four scenarios reflect experiments detailed in the main text: a standard 5-layer network (Fig.~1a), a 5-layer network with naive initialization (Fig.~3), a deeper 10-layer architecture (Fig.~1b, Fig.~4), and a 5-layer network under the feedback alignment scenario (Fig.~6a). 

The results in Fig.~\ref{fig:sup_y_norms} demonstrate that incorporation of Oja's rule consistently maintains activation norms, preventing destabilization, thus improving learning efficiency. This rapid re-scaling plays a key role in ensuring the model performs robustly under low-data conditions.

\subsection{Layer-wise invertibility via transposed weights \label{sec:reverse}}

\begin{figure}[H]
    \centering
    \includegraphics[width=0.49\textwidth]{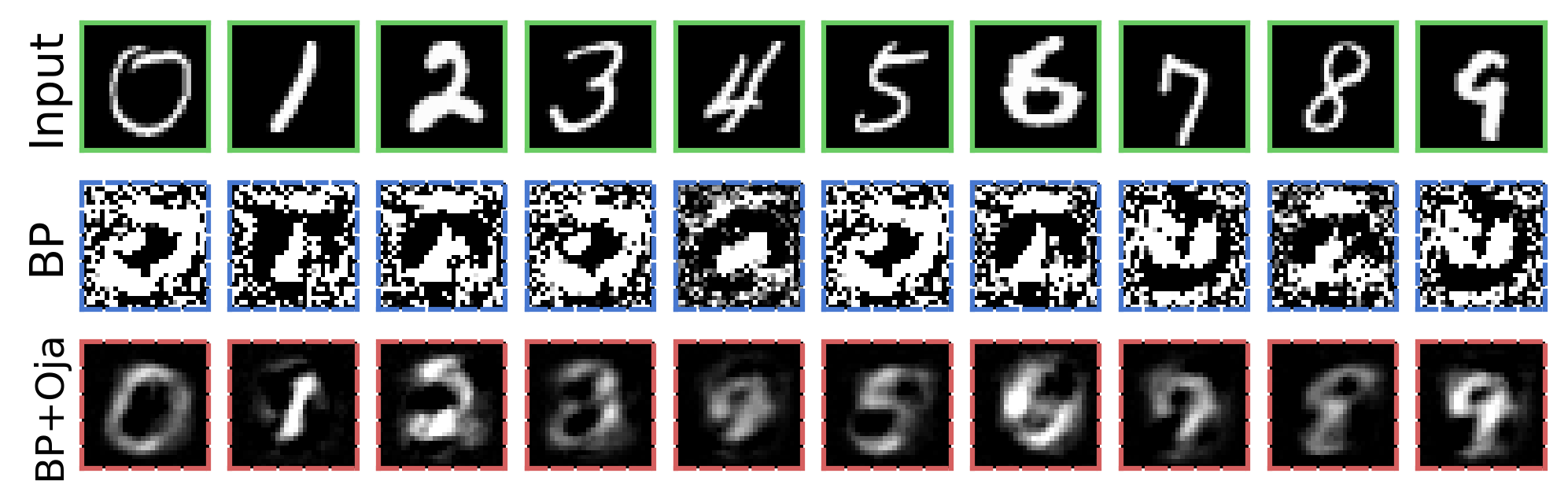}
    \caption{\textbf{Oja's rule promotes layer-wise invertibility.} 
    \textbf{Top row}, original MNIST digits. 
    \textbf{Middle row}, reconstructions from a network trained with backprop, using $\boldsymbol{W}_{\ell-1, \ell}^{\top}$ to invert each layer. 
    \textbf{Bottom row}, reconstructions from a network trained via the hybrid backprop combined with Oja's plasticity. The hybrid model provides significantly clearer images, indicating enhanced invertibility through Oja's rule.}
    \label{fig:bp_oja_transpose}
\end{figure}

In the main article, we assess the preservation of upstream information by reconstructing inputs from the final hidden layer using optimal linear maps $R_{\ell,\ell-1}$ (Fig.~2c; see Methods). Here, we extend this analysis by examining the layer-wise invertibility of the network weights. Formally, we replaced the optimal mappings $\boldsymbol{R}_{\ell,\ell-1}$ with the transposed forward weights $\boldsymbol{W}_{\ell-1, \ell}^{\top}$, thereby estimating:
\begin{equation}
    \hat{\mathbf{y}}_{\ell-1} \;=\; \boldsymbol{W}_{\ell-1, \ell}^{\top}\,\hat{\mathbf{y}}_{\ell},
    \quad
    \text{where}
    \quad
    \hat{\mathbf{y}}_{L-1} = \mathbf{y}_{L-1}
\end{equation}
Remarkably, networks trained using the hybrid rule (equation~(\ref{eq:BPOja})) yielded substantially clearer reconstructions than those trained solely via backprop (Fig.~\ref{fig:bp_oja_transpose}). In the purely backprop-trained networks, transposed reconstructions were illegible, underscoring that Oja's rule enhances the invertibility of the learned weights. 

Since improved invertibility permits the backward pass to recover forward activations on the fly, it obviates the need to store intermediate activations in memory. Hence, combining Oja's rule with backprop can reduce memory overhead by allowing the necessary signals for gradient updates to be reconstructed directly from the final hidden layer.

\subsection{Performance of Sanger's rule in hybrid error-driven learning \label{sec:sanger}}

In the main text, we establish that augmenting both standard backprop and biologically inspired feedback alignment with Oja's rule leads to notable performance gains. Here, we consider Sanger's rule~\cite{sanger1989optimal}, a hierarchical Principal Subspace Analysis (PSA) algorithm, to determine if it provides a comparable benefit. Sanger’s rule removes the upper off-diagonal terms in Oja’s normalization, thereby imposing a strict ordering on principal components.

\begin{equation}
    \mathcal{F}^{\text{Sanger}} 
    \;=\;
    \underbrace{-\theta_1 \,\mathbf{e}_{\ell} \,\mathbf{y}_{\ell-1}^\top}_{\text{backprop term}}
    \;+\;
    \underbrace{\theta_2 (\mathbf{y}_{\ell} \mathbf{y}_{\ell-1}^\top
    \;-\;
    \operatorname{tril}(\mathbf{y}_{\ell} \mathbf{y}_{\ell}^\top)\,\boldsymbol{W}_{\ell-1,\ell})}_{\text{Sanger's term}},
\end{equation}
By fixing the principal subspace basis, this adjustment limits the rotational flexibility granted by Oja’s rule.

Mirroring the approach in the main text, we train a 10-layer network with backprop and a 5-layer network with feedback alignment (see ``Methods: Models''). We employ a meta-learning framework (see ``Methods: Meta-Learning Plasticity'') to tune plasticity parameters for each rule. Each model is meta-trained on 5-class classification tasks over 500 episodes, with every episode including 50 training samples and 10 query samples per class.

\begin{figure}[H]
    \setlength{\unitlength}{0.01\textwidth}
    \begin{picture}(98,31)
        \put(1,29.5){\footnotesize a}
        \put(2,0){\includegraphics[width=0.46\textwidth]{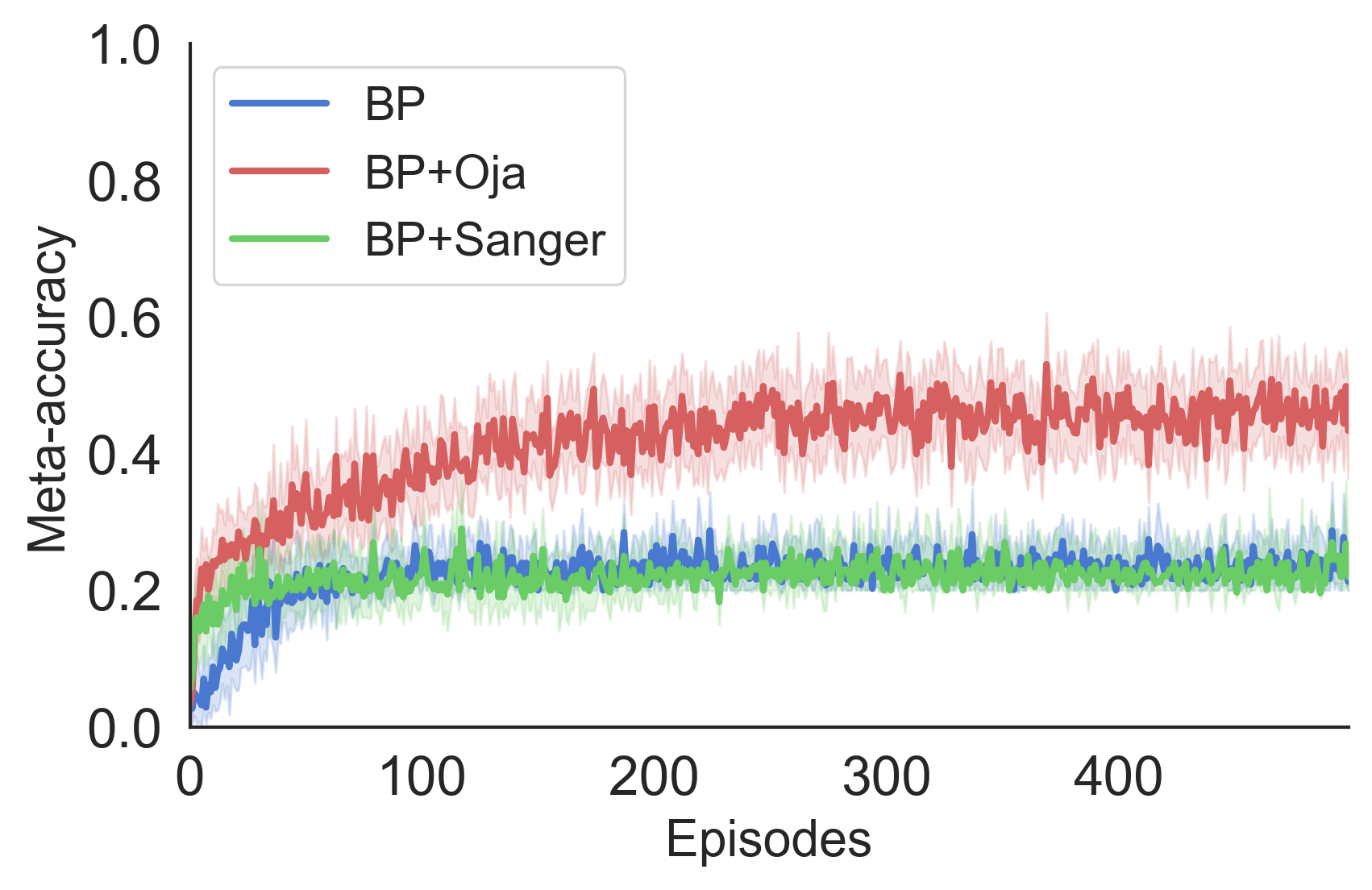}}
        \put(51,29.5){\footnotesize b}
        \put(52,0){\includegraphics[width=0.46\textwidth]{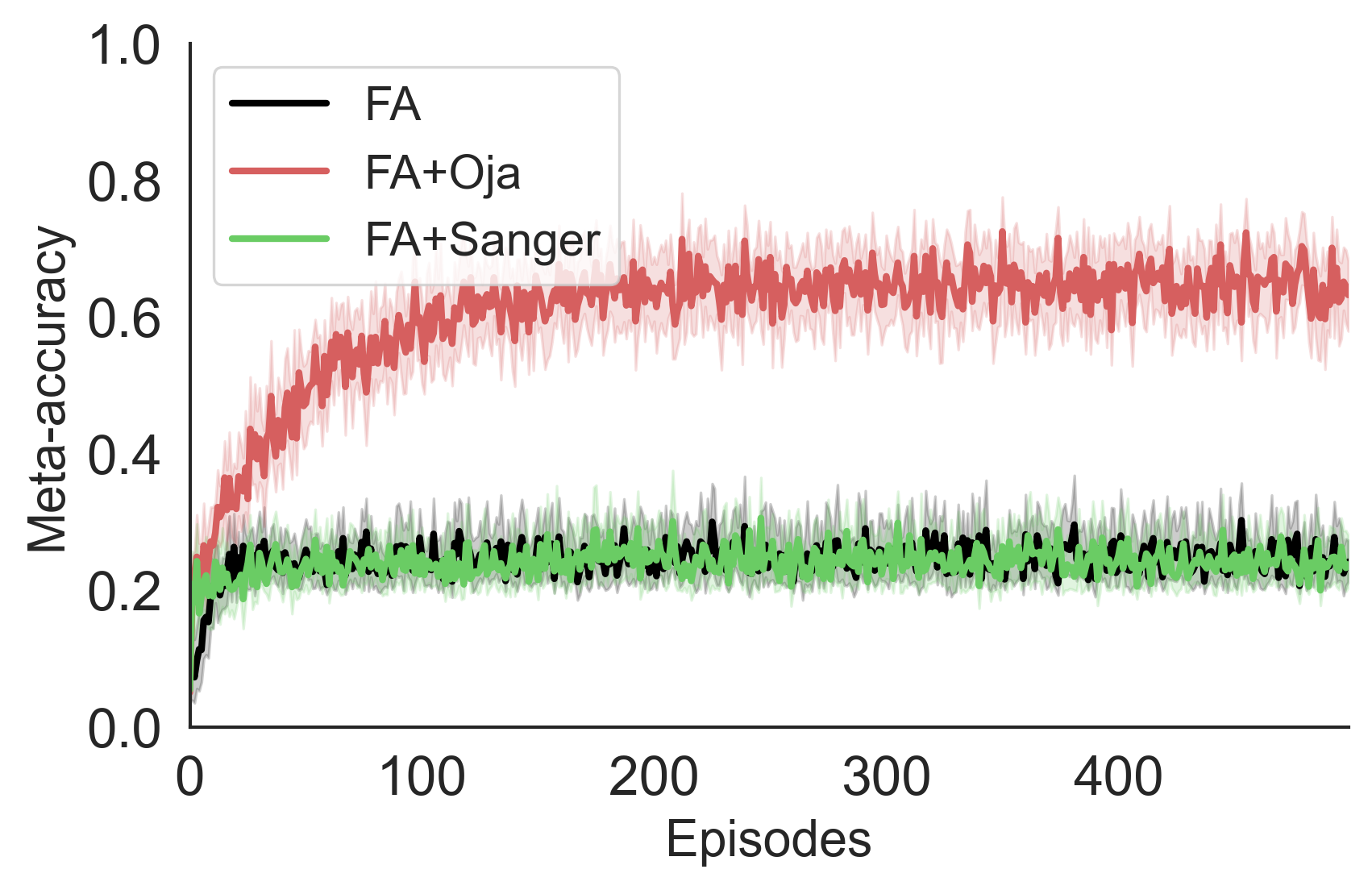}}
    \end{picture}
    \caption{\textbf{Sanger's rule does not improve learning in hybrid settings.} 
    \textbf{a}, Meta-accuracy versus training episodes for a 10-layer network comparing backprop (BP; blue), BP+Oja (red), and BP+Sanger (green). 
    \textbf{b}, Meta-accuracy for a 5-layer network comparing feedback alignment (FA; black), FA+Oja (red), and FA+Sanger (green). Shaded regions indicate $98\%$ confidence intervals across 20 seeds via 500 bootstrap samples.}
    \label{fig:Sanger-Oja}
\end{figure}

Our first experiment incorporates Sanger’s rule into backprop. Figure~\ref{fig:Sanger-Oja} reveals no improvement over backprop alone, in contrast to the substantial gains achieved by Oja’s rule. In a subsequent experiment, we similarly combine Sanger’s rule with random feedback alignment (FA) while preserving the same dataset and task structure. Once again, we observe no benefit relative to FA alone.

We attribute this lack of improvement to the strict convergence requirements of Sanger’s rule, which must align the weights with the true eigenvectors rather than a rotated basis. This constraint diminishes representational flexibility, whereas Oja’s symmetry offers adaptibility under gradient-driven learning.

Overall, these results confirm our main-text conclusion that \emph{symmetric} PSA algorithms, exemplified by Oja’s rule, integrate more naturally with error-driven learning mechanisms than do \emph{hierarchical} approaches such as Sanger’s rule. Further investigation into other symmetric subspace methods could reveal similarly advantageous flexibility for hybrid learning networks.

In summary, these findings reinforce our main-text conclusion that \emph{symmetric} PSA methods, such as Oja’s rule, better complement error-driven processes than \emph{hierarchical} PSA approaches like Sanger’s rule. Although other symmetric techniques may offer comparable flexibility, exploring them remains an open direction for future work.

\end{document}